\documentclass[journal]{IEEEtran}
\usepackage{multirow,pstcol}
\usepackage{epsfig,cite}
\usepackage{amsmath,graphicx}
\usepackage{pstricks}
\usepackage{subfig}
\usepackage[tight,footnotesize]{subfigure}

\begin{document}
\title{A Novel Maneuvering Target Tracking Approach by Stochastic Volatility GARCH Model}

\author{Ehsan~Hajiramezanali,
        Seyyed~Hamed~Fouladi, and Hamidreza~Amindavar
\thanks{Ehsan Hajiramezanali is
with the Department of Electrical Engineering,
Texas A\&M University, College Station, TX, USA, (E-mail: ehsanr@tamu.edu).}
\thanks{Seyyed~Hamed~Fouladi is
with the Department of Electrical Engineering,
Norwegian Institute of Science and Technology, Trondheim, Norway, (E-mail: hamed.fouladi@ntnu.no).}
\thanks{Hamidreza Amindavar is
with the Department of Electrical Engineering,
Amirkabir University of Technology, No. 424, Hafez Ave., Tehran (15914), Iran,
(Phone: (+98) 21-6454-3332, Fax: (+98) 21-6640-6469, E-mail:
hamidami@aut.ac.ir).}
}

\maketitle

\begin{abstract}
%\boldmath
In this paper, we introduce a new single model maneuvering target
tracking approach using stochastic differential equation (SDE) based on GARCH
volatility. The traditional input estimation (IE) techniques
assume constant acceleration level which do not  cover all the
possible acceleration quintessence. In contrast, the multiple model
(MM) algorithms that take care of some
IE's shortcomings, are sensitive to the transition probability matrices.
In this paper, an innovative model is proposed to overcome these drawbacks
by using a new generalized dynamic modeling of acceleration and a
Bayesian filter. We utilize  SDE to model Markovian jump
acceleration of a maneuvering target through GARCH process
as the SDE volatility.
In the proposed scheme, the original state and stochastic volatility (SV)
are estimated simultaneously by a bootstrap particle filter (PF).
We introduce the bootstrap resampling to obtain the
statistical properties of a GARCH density. Due to the heavy-tailed
nature of the GARCH distribution, the bootstrap PF  is more effective in
the presence of large errors that can occur in the state equation. We
show analytically that the target tracking performance  is improved by considering
GARCH acceleration model. Finally, the effectiveness and capabilities
of our proposed strategy (PF-AR-GARCH) are demonstrated
and validated through simulation studies.

\end{abstract}

\begin{IEEEkeywords}
Maneuvering target tracking, stochastic volatility, SDE, GARCH process, bootstrap particle filter.
\end{IEEEkeywords}

\IEEEpeerreviewmaketitle

\section{Introduction}
\IEEEPARstart{M}{aneuvering} target tracking has attracted a great
deal of attention in recent years, due to the significant importance
to a wide range of civilian applications such as intelligent
transportation systems, air traffic control, surveillance, indoor
tracking, cellular radio network and bioinformatics
\cite{Mobile,Indoor,dependent,hajiramezanali2018scalable}. However, there are still many
challenges that make this issue difficult. The difficulty in
tracking a maneuvering target in the presence of false
measurements arises from the indirectly observed acceleration and
the uncertainty in modeling a maneuver in state equation of  the
target \cite{Em}.
The behavior of the moving targets (objects) are governed by physical laws which
can be expressed as mathematical equations. These equations
represent how quantities such as position, speed and acceleration, etc.,
change from their current values (at the present time).
The best target position prediction models are global models that solve the
mathematical equations governing the behavior of the target(s) at
every point.

Significant research attempts have been allocated
to the problem of maneuvering target tracking.
In the history of development of the target tracking
techniques, single model based Kalman filtering
came into existence first,
but when the target maneuver occurs, its performance
is often seriously degraded.
As a first attempt to solve this problem,
Singer \cite{Singer} proposed a typical whole statistic model where
the second-order statistics of acceleration is the same as
a first-order Markov process.
Although a tracking filter with Singer's model shows good
performance for the target with a low maneuver,
its performance is rapidly corrupted in case of high
maneuvering and constant velocity targets.
Moreover, some a priori statistical descriptions must
be given to the maneuver process, that often
requires more knowledge about the object than what is
normally available \cite{VD,IE,MIE,Sur,novel,CT,TR}.

In recent years, researchers devoted a great deal of attention to
the decision-based methods, which detect
the maneuver and then cope with it effectively.
Examples of this approach including the IE
techniques \cite{IE,IE2,IE3}, the variable dimension (VD) filter \cite{VD}, etc.
In addition to the basic filtering computation,
these techniques require a great deal of effort to detect
the target abrupt accelerations \cite{digital,Neural}.
In the IE scheme, which is widely accepted
as one of the most effective decision-based methods,
the magnitude of the acceleration is identified in a least
square format when a maneuver is detected . Although this method is non-parametric,
the detection algorithm requires a significant
amount of computation and memory \cite{VD}. Therefore, a
large delay exists in this approach \cite{MIE,MIE2}.
Furthermore, this technique shows poor performance
in case of low maneuvering target.
Although the modified
input estimation (MIE) \cite{MIE,khal,MIE2} demonstrates
a reasonable performance in tracking low-maneuvering targets,
its performance leads to a serious degradation
in the presence of high maneuvers.

On the other hand, the motion of a target can be
described with MM techniques such as
the generalized pseudo-Bayesian (GPB) method \cite{GPB},
the interacting MM (IMM) method
\cite{IMMbook,IMMcontrol,IMM2001,IMM2003,novel,IMMground,MTT}, and etc.,
which describe the motion of a target using
multiple sub-filters.
Among MM algorithms, the
IMM method is the most common one.
However, in the IMM estimators, the possible motion
models and transition probability matrices are assumed to be known \cite{digital,novel}.
In practice, the dynamics are hard to break up into several
different motion models and the model transition probabilities
are difficult to obtain \cite{novel,MMM,minimal}.
In addition, with the increase in the number of models, the
computational cost will also increase significantly, which
seriously affects the real-time performance for tracking a
maneuvering target.
Nevertheless,
they have the common disadvantages: (a) an unnecessary amount
of computations when the target is maneuvering and (b) the
potential loss of accuracy due to an over modeling at these maneuvering
times \cite{Novelex}.

On the other hand, a maneuver can be related to time series
that govern the state space model. All the mentioned approaches \cite{IE,IE2,IE3,MIE,MMM}
constrain the process noise sequence of the state model to be Gaussian.
The Gaussian
assumption on the process noise is not universal
because a target's motion can be
affected by the combination of small perturbations
due to the air turbulence and  controller induced
changes to the speed and course \cite{levy,digital}. Moreover,
the state space noise for  a maneuvering target can be a Markovian
jump process which is unlikely to follow a stationary
Gaussian probability density function.
Through this paper, we have proven analytically, the impact of
maneuver in the process noise. Moreover,
we have shown that the impact of maneuver
on the process noise of the state model,
can be effectively modeled as the GARCH process.

In this paper, we expand on the target tracking based on
volatility modeling in conjunction with SDE \cite{ISSPA, SSP, ETRI}.
The proposed GARCH method is suitable for
both maneuvering and non-maneuvering target motions.
The SV, i.e. time varying variance, causes the model error covariance
to be time varying. However, the time-varying conditional covariance is
considered for the state equation which can better describe the maneuver.
To take the advantage of SV in (non)-maneuvering target,
we calculated the stochastic It$\bar{\mbox{o}}$ integral to obtain the covariance matrix.
The advantages of the proposed method are two fold. First,
the state estimation accuracy of our method
outperforms in comparison with IE, MIE and IMM approaches especially in
high maneuvering targets (jumpy acceleration maneuver).
Secondly, the computational complexity of the PF-AR-GARCH scheme
is low compared to the PF-MM algorithms. We show analytically that the target tracking
performance, by considering GARCH acceleration model, is improved,
especially abrupt changes in acceleration tracking.
In the proposed scheme, the original state and SV
are estimated simultaneously with a particle filter
and unlike the decision based algorithms such as IE,
the maneuver detection procedure is eliminated.
Finally, the effectiveness and capabilities of our proposed strategy are demonstrated
and validated through simulation studies.
Simulation tests show that the proposed PF-AR-GARCH technique is superior to the traditional MIE
and IMM methods in estimation accuracy especially in
the velocity and acceleration estimations.
These large improvements
in velocity and acceleration estimations are particularly
useful in some tactical applications such as threat evaluation,
the computation of a hostile missile's flight time, and etc. \cite{Freq}.

The paper is organized as follows.
In section II, the GARCH process is introduced and the
suitability of the GARCH for modeling the acceleration of
targets is provided. Section III is allocated to describe the new
problem formulation and proposed the target equation motion based on GARCH model.
Section IV is appropriated to bootstrap filtering for non-linear
and non-Gaussian GARCH target model. The complexity of the proposed method is evaluated in section V.
Simulation results
with various scenarios are discussed in section VI.  At the end, concluding remarks
are provided in section VI.

\section{Problem statement}
In this section, we study whether GARCH modeling provides a flexible and
appropriate tool for modeling the acceleration of (non)-maneuvering
targets. To examine the suitability of the GARCH for modeling the
acceleration of targets, first,
the GARCH process is introduced and its
properties are discussed in subsection $A$.
In the second subsection, we show analytically that the
stochastic behavior of autoregression coefficients of traditional
acceleration models leads to the conditional heteroscedasticity.
At the end of this section,
we describe that SV in SDE captures
the features of targets acceleration.
Moreover, in order to select the true estimation scheme of SV,
we explain why GARCH model is suitable.

\subsection{GARCH Model}
The traditional time series focus on modeling the conditional first moment under an
assumption of constant variance.
To generalize the implausible
assumption of constant variance, a class of stochastic processes
called Autoregressive Conditional Heteroscedastic (ARCH)
processes were introduced by Engle
in \cite{En}. These processes allow the
conditional variance to change over time as a function of past
innovations.
A more general
class of stochastic processes, GARCH, was introduced by Bollerslev in
\cite{GARCH} with more
flexible lag structure.

The AR-GARCH process, which is a filtered version of a GARCH process with an
all-pole filter \cite{AR,hajiramezanali2018differential}, is a suitable model for capturing the statistical properties of acceleration
because the accelerations of targets can be correlated with time. Therefore,
the linear correlation of acceleration can be modeled by autoregression coefficients.
The AR-GARCH model describes time-varying variance
of process and correlation of process with time simultaneously. $a(k) $
follows a pure AR($r$)-GARCH($p,q$) model if

\begin{small}
	\begin{eqnarray}
	\label{model1} a(k)  &=& \sum_{n = 1}^r {b(n) a(k-n)  +
		z(k) },\quad z(k)  = \sqrt {h(k) } \varepsilon (k) \\
	\label{model2} h(k)  &=& \alpha_0  + \sum_{i=1}^{p} {\alpha _i z^2(k-i)}  + \sum_{i=1}^{q} {\beta_i h(k-i)},
	\end{eqnarray}
\end{small}
where $\alpha _0\  > 0,$  $\alpha_i , \beta_i  \geq 0,$ $i$ is the time
index, $b(n) $ and $r$ in (\ref{model1}) are the parameters
and the order of AR part respectively, $\varepsilon (k) $ is
a sequence of zero-mean iid random variables with unity variance,
and $h(k) $ is the conditional variance of
$z(k) $. In practice, $\varepsilon(k) $ is often assumed to be
independent Gaussian random variables.
According to \cite{GARCH}, the GARCH process,
defined in (\ref{model1}) and (\ref{model2}),
is wide sense stationary if and only if $\alpha_i + \beta_i  \leq 1$.
From (\ref{model2}) it is obvious that
at each time, both the neighbouring sample variance and the
neighbouring conditional variance play a role in the current
variance.

A characteristic feature of GARCH series is the volatility clustering,
where the periods of high and low volatility occur in the data.
Typically, the changes between periods of low, medium, and high
volatility do not exhibit any systematic patterns and seem to be the best
model as occur randomly \cite{Mag}. On the other hand,
the target motions can have different values for the mean of the
continuous time acceleration during each segment of their
trajectories in the same way as the volatility characteristics of
a GARCH series. In the other words, GARCH process follows a
fundamental philosophy about the next status of target in the
maneuvering target tracking application. In the proposed GARCH
model, GARCH process considers that the target is in the
(non)-maneuver situations with higher probability if the target is in
the (non)-maneuver status at the previous times. The
GARCH(1,1)\footnote{The simplest GARCH(1,1) is suitable for all
practical scenarios \cite{B2,GARCH11}.} indicates that the predicted
variance rate is based on the most recent observation of the
squared average and the most recent estimate of the variance rate.
Along with the ability to model the volatility evolution, the
GARCH model also assigns weights that decrease exponentially with
respect to the past observations in the data. Therefore, more
recent jumps in acceleration have more impact on the state space
model. This could yield an advantage for using GARCH modeling
of $h(k)$ in the maneuvering target tracking over Heston modeling, or
Barndorff-Nielsen and Shephard model. Furthermore, because of the
effects of the recent samples in diffusion, this model also
encompasses the ``{\em black swan}" events which refers to a
sudden large, unexpected movement in the process. In the proposed
acceleration model, a large value of $|a_{t-1}|$, which is an
indication of high volatility of $a_{t-1}$, increases $h_t$, the
volatility of $a_t$. In this paper, we utilized the GARCH model in
tracking simulations and it has been shown to give a suitable
representation of the target’s maneuver and non-maneuver
trajectories.

Using the results of Bollerslev in
\cite{GARCH}, it is possible to derive %the following expression of
the kurtosis of GARCH(1,1) to determine.
\begin{eqnarray}
\nonumber
\kappa&=&\left(\mbox{E}\left(\varepsilon^4(k) \right)-3\mbox{E}\left(\varepsilon^2(k)\right)^2\right)\mbox{E}\left(\varepsilon^2(k)\right)^{-2}
\\ &=&6\alpha_1^2\left(1-\beta_1^2-2\alpha_1\beta_1-3\alpha_1^2\right)^{-1}
\end{eqnarray}
where $\kappa$ is the kurtosis of residual $\varepsilon(k)$. If kurtosis is greater than 0 ($\kappa \ge 0$) by the assumption of
moment conditions for GARCH(1,1) (Fig. 1 in \cite{GARCH}),
which can be greater than the kurtosis of Gaussian distribution, then GARCH(1,1) is found to successfully model a heavy-tail random process.
Because the abrupt changes in acceleration are translated into heavy-tail
volatilities of acceleration, hence, we expect $\kappa\geq 0,$ and
GARCH(1,1) suitably model target acceleration.
Therefore, the proposed target acceleration modeling is provided as an AR(1)-GARCH(1,1) process
in the following.

\subsection{Conditional Heteroscedasticity of Acceleration}
Here, we demonstrate analytically the suitability of the GARCH model for the modeling of
target acceleration. The target acceleration, $a$, is correlated in time, i.e. acceleration
in time instant $k$, $a(k),$ can be predicted by the previous value of acceleration in time
instant $k-1$, $a(k-1),$ with a random prediction error $w(k)$.
According to the practical example described in \cite{Singer},
acceleration correlation in a slow turn often
rises for up to one
minute, acceleration correlation in evasive maneuvers will be provided
for the periods between ten to thirty seconds, and acceleration correlation in atmospheric turbulence
will be provided for one or two seconds.
In \cite{Singer},
the discrete acceleration model is expressed as below:
\begin{eqnarray} \label{AccSinger}
\nonumber a(k)&=&e^{-\mu T} a(k-1)+w(k), \\
w(k)&=&\sigma_m^2 z(k)
\end{eqnarray}
where $\sigma_m^2$ is the conditional variance of $a(k)$,
i.e. $\mbox{Var} \left. \left\{a(k) \right| a(k-1)\right\}= \sigma^2_m,$
and $z(k)$ is an i.i.d Gaussian
random process.
$\mu$ is the reciprocal of the acceleration time constant.
For instance, $\mu$ is approximated $1/60$, $1/20$ and $1$ for
a slow turn, in an evasive maneuver and atmospheric turbulence, respectively.
It is clear that the value of $e^{-\mu T}$ cannot be considered as a constant
value under the realistic circumstances. Therefore, we suppose that the autoregression coefficient
$e^{-\mu T}$ in (\ref{AccSinger}) is adjusted to behave as a stochastic process $\zeta(k).$
Generally speaking, the stochastic process $\zeta(k)$ could exhibit very small, very large, or no
change at all, in time. Accordingly, equation (\ref{AccSinger})
becomes:
\begin{equation} \label{AccSinger1}
a(k)=\zeta(k) a(k-1)+w(k).
\end{equation}
In appendix~A, given the noisy acceleration record $a(k)$
for  $0 < k < k-1$ in (\ref{AccSinger1}),
we prove that the stochastic property of autoregression coefficient of acceleration
in (\ref{AccSinger1}), $\zeta(k),$
leads to the conditional heteroscedasticity.
Then, in \cite{TSP}, we have analyzed that
even small changes in $\zeta(k)$,
result in large changes in the conditional variance of
$a(k)$. The conditional variance of $a(k)$, in terms of the conditional variance, $\zeta(k)$ is expressed as:
\begin{equation} \label{AccSinger2}
\mbox{Var} \left. \left\{a(k) \right| a(k-1)\right\}=\sigma_{\zeta}^2 a^2(k-1)+\sigma^2_m
\end{equation}
where $\sigma_{\zeta}^2$ is the
conditional variance of $\zeta(k)$ which can be a constant or
time-varying in time. Therefore, it is not a realistic assumption
on the conditional variance $a(k)$ to be constant in time, seen
from (\ref{AccSinger2}).
Then, $\mbox{Var} \left. \left\{a(k-1)
\right| a(k-2)\right\}$ determines the present conditional
variance $\mbox{Var} \left. \left\{a(k) \right| a(k-1)\right\}$
seen in the following derivations.
Referring to (\ref{AccSinger2}) and using (\ref{AccSinger1}),
the conditional variance can be rewritten as:

\begin{small}
\begin{eqnarray} \label{AccSinger3} \nonumber
\mbox{Var} \left. \left\{a(k) \right| a(k-1)\right\} &=& \sigma_{\zeta}^2 \left(\zeta(k-1) a(k-2)+w(k-1) \right)^2+\sigma^2_m\\
\nonumber
&=& \sigma_{\zeta}^2 \zeta^2(k-1) a^2(k-2) + \sigma_{\zeta}^2 w^2(k-1)\\
&+& 2 \sigma_{\zeta}^2 \zeta(k-1) a(k-2) w(k-1) + \sigma_m^2.
\end{eqnarray}
\end{small}
We can calculate $a^2(k-2)$ according to (\ref{AccSinger2}) as below:
\begin{eqnarray} \label{AccSinger4}
a^2(k-2) = \frac{1}{\sigma^2_\zeta}\left( \mbox{Var} \left. \left\{a(k-1) \right| a(k-2)\right\} - \sigma^2_m \right).
\end{eqnarray}
By inserting (\ref{AccSinger4}) in (\ref{AccSinger3}), the conditional variance $\mbox{Var} \left. \left\{a(k) \right| a(k-1)\right\}$
is expressed as:

\begin{small}
\begin{eqnarray} \label{AccSinger5} \nonumber
&&\mbox{Var} \left. \left\{a(k) \right| a(k-1)\right\} = \\\nonumber&& \sigma_{\zeta}^2 \zeta^2(k-1) \left( \frac{1}{\sigma^2_\zeta}\left( \mbox{Var} \left. \left\{a(k-1) \right| a(k-2)\right\} - \sigma^2_m \right)\right) +\\ \nonumber
&& 2 \sigma_{\zeta}^2 \zeta(k-1) a(k-2) w(k-1) + \sigma_m^2 + \sigma_{\zeta}^2 w^2(k-1) =\\ \nonumber
&& \zeta^2(k-1) \mbox{Var} \left. \left\{a(k-1) \right| a(k-2)\right\} - \zeta^2(k-1) \sigma^2_m + \\
&& 2 \sigma_{\zeta}^2 \zeta(k-1) a(k-2) w(k-1) + \sigma_m^2 + \sigma_{\zeta}^2 w^2(k-1).
\end{eqnarray}
\end{small}
Therefore, we conclude that
conditional variance (\ref{AccSinger5}) can be effectively described by GARCH model,
in subsection $A$.
In the SDE literatures, the time varying conditional variance
in (\ref{AccSinger2}-\ref{AccSinger5}) is named as stochastic volatility. Therefore, we consider
$h(k) = \mbox{Var} \left. \left\{a(k) \right| a(k-1)\right\}$ in the following
where $h(k)$ is the stochastic acceleration volatility.

\subsection{Stochastic Volatility Estimation}
The methodology to be presented here is based on the fact that the
target acceleration, $a_t$, can be termed as the target maneuver
variable. In order to increase the acceleration modeling
capabilities not affected by sampling time, we present continuous
acceleration  instead of a discrete modeling. In the present work,
the acceleration volatility and hence, the amount of target
maneuvering, is correlated in time; namely, if a target is
accelerating with large (small) variance at time $t$, it is likely
to be accelerating with large (small) variance at time $t + \tau$.
In the same way, time series typically exhibit time-varying
volatilities; volatility is an index of time-varying variance
and/or correlation. Higher volatilities in the state space noise
demonstrate an increased chance of maneuvering motion. Therefore,
the models of time varying volatilities and correlations are essential
for the maneuver management.

We model acceleration in terms of It$\bar{\mbox{o}}$ stochastic differential equation as in
\begin{equation}
\label{model4} da_t  = \mu \,a_t\, dt + \sqrt {h_t }\, dW_t
\end{equation}
where $\mu $ is the constant drift of the target acceleration $a_t $,
$h_t $ is the stochastic acceleration volatility, and $dW_t$ denotes a
white Brownian motion.
In (\ref{AccSinger1})-(\ref{AccSinger5}), we proved that
the non-blind acceleration modeling by Singer \cite{Singer} can be
formulated to encompass larger class
of accelerations exhibited by a target if we
consider GARCH modeling of acceleration via an SDE in (\ref{model4}).

The key problem of the maneuvering target tracking in the proposed
SDE model is to estimate  SV of the acceleration.
Estimation of SV is a crucial step in the present work.
Considering $h_t$ as the particular SV for the model under study in (\ref{model4}), it is natural
to model the variance function, $h_t$, as another Brownian motion in the continuous time.
For such an aim, three well-known models are described below:
\begin{enumerate}
\item Heston model: The popular Heston model is a commonly used SV model,
in which,
the variance in (\ref{model4}), $h_t$, obeys the following
mean-reverting SDE
\begin{equation}
\label{Heston} dh_t  = \theta \left( {\omega  - h_t } \right)dt + \xi \sqrt{h_t} dB_t.
\end{equation}
Here, $\omega $ is the
long-term mean volatility, $\theta $ is the rate at which the
volatility reverts toward its long-term mean, $\xi $ is the
conditional standard deviation of the volatility process, and $dB_t $ is another
standard Gaussian Brownian motion that is correlated with $dW_t $ of (\ref{model4}) with constant
correlation factor $\rho$.
Equation (\ref{Heston}) is known as the Cox-Ingersoll-Ross (CIR) process in the financial literature \cite{Cox}.
\item GARCH model: In this model, the variance function of the
target acceleration, $h_t$, in (\ref{model4}) is estimated
using the GARCH model described by another SDE
\begin{equation}
\label{GARCH} dh_t  = \theta \left( {\omega  - h_t } \right)dt + \xi h_t dB_t.
\end{equation}
The GARCH model assumes that the randomness of the
variance process varies with the diffusion \cite{Mag}, opposed to the Heston model where it varies with the
square root of the diffusion.
\item Barndorff-Nielsen and Shephard model: Another approach is the SV model of
Barndorff-Nielsen and Shephard \cite{OU} in which the volatility process, $h_t$,
is an Ornstein-Uhlenbeck process driven by a nondecreasing L$\acute{\mbox{e}}$vy process
\begin{equation}
\label{OU} dh_t  = -\lambda h_t dt + dz(\lambda t), \quad \lambda>0
\end{equation}
where $z$, with $z(0) = 0$, is a homogeneous L$\acute{\mbox{e}}$vy process, i.e. a process with independent and
stationary increments.
\end{enumerate}
Before using the GARCH model for the estimation of SV, it is essential
to discuss the plausibility of GARCH modeling of $h_t$ for a
special maneuvering target tracking application. The validity of
this claim is proved by three reasons. (a) Our ultimate goal is to
track $h_t$ (a Markovian jumpy process). As we demonstrated in the
subsection A, the stochastic volatility $h_t$, which is
described as conditional variance, causes conditional
heteroscedasticity of acceleration. Therefore, GARCH process is a
proper model for $h_t$ in this application because of its
fundamental compatibility for heteroscedastic processes.
(b) GARCH models account for the volatility clustering;
i.e., large changes tend to follow large changes and small
changes tend to follow small ones, are compatible to a large
extent to the maneuvering target. Naturally, we can observe
the same characteristic for target motion, i.e. motions can
have low and high accelerations all the way along a trajectory
in a period of time.
(c) As elaborated in the introduction, the process noise is not Gaussian universally \cite{levy,digital}.
Therefore, the use of Gaussian system noise causes
blunt estimation to abrupt changes of the state.
To overcome this
problem, we propose the use of uni-modal heavy-tailed non-Gaussian
distribution for innovation process. GARCH model is capable of taking
into account this characteristic of noise, namely heavy-tailed distribution.

The difference in Heston model in
comparison to GARCH approach is only in the
square root of volatility. This square root
reduces the entropy of volatility, hence, GARCH could encompass more possible
target accelerations. On the  other hand,
in our opinion, Barndorff-Nielsen and Shephard model
is more general than GARCH modeling for the volatility. This is
because L$\acute{\mbox{e}}$vy processes are more general
than GARCH processes \cite{OU,SDE}, however, the
amount of computation required for such modeling is a lot more than GARCH approach.
Based on the comparison between the features of the maneuverity in the
acceleration and the structures of the GARCH model, one can conclude that
the GARCH model is the most appropriate one for this application.
Therefore, we estimate the stochastic volatility, $h_t$, of the target acceleration in
(\ref{model4}) using the GARCH model illustrated in (\ref{GARCH}).

\section{problem formulation}
Realistic tracking applications require a suitable target model which is simple enough
for the execution in tactical systems. On the other hand, the computational complexity of the target model
should be at a bounty level not compromising the accuracy of tracking.
Because of its real-time limitations,
the target model offered here accounts for this
objective in a way that it is simple and provides a proper representation
of the maneuvering target behavior.
Due to the continuous nature of an actual target motion,
it is often more appropriate to use the
continuous target dynamic models for most tracking
problems \cite{Sur,Singer,hajiramezanali2018differential}. In other words,
target motions should not depend on how
and when the samples are taken, which is often the case,
however, for a discrete-time model.
In this paper, the maneuvering target equations are presented
by continuous time model and are then expressed in
discrete time using the standard discretization procedure,
thereby providing an accurate statistical representation of the true target behavior.

The maneuver model considered here is a new approach which detects
the existence of target maneuvers and directly estimates the
magnitude of the unknown parameters. In the following, the
mathematical formulation of the problem of tracking maneuvering
targets  using  SDE algorithm will be discussed.

\subsection{Dynamic Equations of Target Motion}
To define the problem of tracking, consider a dynamic system
represented by the state sequence, whose temporal evolution is
provided by two n-dimensional SDE
\begin{small}
\begin{eqnarray}
\label{model8} \!\!\!\!\!\!\!\!\!\!\!\!\!\!\!&& d\mbox{\boldmath
$X$}(t) = \mbox{\boldmath $F$}\mbox{\boldmath $X$}(t) dt +
\mbox{\boldmath $G$} \mbox{\boldmath $H$}^{\frac{1}{2}}(t)
d\mbox{\boldmath $W$}(t) \\ \nonumber
\!\!\!\!\!\!\!\!\!\!\!\!\!\!\!&& d\mbox{\boldmath $H$}(t) = \\ \label{model9}
\!\!\!\!\!\!\!\!\!\!\!\!\!\!\!&&
\left[
{\begin{array}{*{20}c}
   %{\theta_x \left( {\omega_x  - h_x(t)}\right) + \rho_x dB_x(t)} & \!\!\!\!\!\!\!\!\!\!\!\!0 \\
   {\theta_x \left( {\omega_x  - h_x(t)}\right)dt + \rho_x dB_x(t)} & \!\!\! 0 \\
%   0 & \!\!\!\!\!\!\!\!\!\!\!\!{\theta_y \left( {\omega_y  - h_y(t)}\right) + \rho_y dB_y(t)} \\
   0 & \!\!\! {\theta_y \left( {\omega_y  - h_y(t)}\right)dt + \rho_y dB_y(t)} \\
\end{array}} \right]
\end{eqnarray}
\end{small}
where $t$ denotes the continuous time index,
$\omega_x$ and $\omega_y$ are the
long-term mean volatilities, and $\theta_x$ and $\theta_y$ are the rates at which the
volatilities revert toward their long-term means of direction x and y, respectively,
and
\begin{equation}
\label{pi} \mbox{\boldmath $X$}(t)=\left[\mbox{\boldmath $x$}(t)^T\quad \mbox{\boldmath $v$}(t)^T\quad
\mbox{\boldmath $a$}(t)^T\right]^T
\end{equation}
is a six-dimensional position-velocity-acceleration
parameter vector that $\mbox{\boldmath $x$}(t)=[x(t)\quad y(t)]^T,
\,\mbox{\boldmath $v$}(t)=[v_x(t)\quad v_y(t)]^T$, and
$\mbox{\boldmath $a$}(t)=[a_x(t)\quad a_y(t)]^T$ are the
position, the velocity and the acceleration of the target respectively.
$\mbox{\boldmath $F$}$ is a state transition matrix
\[
\label{model011} \mbox{\boldmath $F$}  = \left[
{\begin{array}{*{20}c}
   0 &\:\: 0 &\:\: 1 &\:\: 0 &\:\: 0 &\:\: 0  \\
   0 &\:\: 0 &\:\: 0 &\:\: 1 &\:\: 0 &\:\: 0  \\
   0 &\:\: 0 &\:\: 0 &\:\: 0 &\:\: 1 &\:\: 0  \\
   0 &\:\: 0 &\:\: 0 &\:\: 0 &\:\: 0 &\:\: 1  \\
   0 &\:\: 0 &\:\: 0 &\:\: 0 &\:\: { - \mu } &\:\: 0  \\
   0 &\:\: 0 &\:\: 0 &\:\: 0 &\:\: 0 &\:\: { - \mu }  \\
\end{array}} \right],
\]
$\mbox{\boldmath $G$}=[0 \:\: 0; 0 \:\: 0; 0 \:\: 0;
0 \:\: 0; 1 \:\: 0; 0 \:\: 1]^T$ is a
matrix for the addition of system noise,
$d\mbox{\boldmath $W$}(t)$ is considered as a 2-dimensional
vector of independent Brownian motion as below:
\begin{equation}
\label{dW} d\mbox{\boldmath $W$}(t) =
\left[ {\begin{array}{*{20}{c}}
{dW_x(t)}\\[.1cm]
{dW_y(t)}
\end{array}} \right],
\end{equation}
and $dB_x(t) $ and $dB_y(t)$ are other
standard Gaussian Brownian motion processes that are correlated with $dW_x(t) $ and $dW_y(t)$ of (\ref{dW}) with constant
correlation factors $\rho_x$ and $\rho_y$, respectively.

\subsection{Discrete Time Equations of Motion}

Since the volatility is independent of the states in the equation
(\ref{model9}), the proposed SDE model for the maneuvering target is
linear in the narrow sense \cite{SDE}.
Therefore, $\mbox{\boldmath $X$}$ can be expressed as the solution
of the following stochastic integral equation \cite{SDE}:

\begin{small}
\begin{equation}
\label{model13} \mbox{\boldmath $X$}(t) = \exp(\mbox{\boldmath
$F$}t)\mbox{\boldmath $X$}(0) + \int_0^t \exp(\mbox{\boldmath
$F$}(t-s))\mbox{\boldmath $G$}\sqrt {h(s)} dW(s)
\end{equation}
\end{small}
where the integral is  called an It$\bar{\mbox{o}}$ integral and
$\mbox{\boldmath $\Phi$}(t) = \exp(\mbox{\boldmath$F$}t)$
is the target state transition matrix.
The appropriate discrete time target equation of motion for the tracking
problems is given by:
\begin{equation}
\label{model19} {\mbox{\boldmath $X$}(k + 1)} = {\mbox{\boldmath $\Phi$}
}(T,\mu){\mbox{\boldmath $X$}(k)} + {\mbox{\boldmath $U$}(k)}
\end{equation}
where ${\mbox{\boldmath $\Phi$}
}(T,\mu)$ can be calculated from $\mbox{\boldmath $\Phi$}(t)$ by \cite{Singer, Optimal} and ${\mbox{\boldmath $U$}(k)}$ is discretized version of
${\mbox{\boldmath $U$}(t)}$ which according to
(\ref{model13}), is expressed below:
\begin{equation}
\label{U} \mbox{\boldmath $U$}(k) = \int_{kT}^{kT+T} \exp(\mbox{\boldmath
$F$}(kT+T-s)) \mbox{\boldmath $G$} \mbox{\boldmath $H$}^{\frac{1}{2}}(s) d\mbox{\boldmath $W$}(s)
\end{equation}
where stochastic volatility matrix, $\mbox{\boldmath $H$}(s)$, is
\begin{equation}
\label{H} \mbox{\boldmath $H$}(s) =
\left[ {\begin{array}{*{20}{c}}
{h_x(s)} & {0}\\
{0}     &  {h_y(s)}
\end{array}} \right].
\end{equation}

As mentioned above, the second term of right side of
(\ref{model13}) is an It$\bar{\mbox{o}}$ integral,
in terms of conventional integral of Riemann's type, this integral cannot be calculated
due to
the irregularity of the noise.
In It$\bar{\mbox{o}}$ formula, the variation of integration, $dW$,
is a stochastic process, hence, there is no  general explicit
solution for an It$\bar{\mbox{o}}$ integral in the practical
applications. The filtering part does not require the discrete
version of It$\bar{\mbox{o}}$ integral directly. The state
equation noise covariance is just derived from (\ref{Qn}),
therefore, it is  directly suitable for filtering. Thus, it is
plausible to introduce a system noise covariance using
It$\bar{\mbox{o}}$ Integral properties. The two first moments of
$\mbox{\boldmath $U$}(k)$ are derived in appendix B for the target
tracking.

Moreover, the forward Euler discretization can be
used to approximate the SV of 2-dimensional stochastic
differential equation
(\ref{model9}) on a discrete time grid.
Let $\left[ {0 = {t_0} <
{t_1} <  \cdots  < {t_M} = T} \right]$ is considered as a partition of time
interval with $M$ equal segments.
The elements of the vector $\mbox{\boldmath $h$}(k) = diag(\mbox{\boldmath $H$}(k))$
are the discrete version of non-zero elements of SV in (\ref{H}) and have the forms:
\begin{eqnarray}
\nonumber
\mbox{\boldmath $h$}(k) &=&
\left[ {\begin{array}{*{20}{c}}
{h_x(k)}\\
{h_y(k)}
\end{array}} \right] \\ \label{eq_H2} &=&
\left[ {\begin{array}{*{20}{c}}
{{\alpha_{0x} + \alpha_{1x} {h_x(k-1)}} + \beta_{1x} {\mathop{z_x^2}\nolimits}(k-1)}\\
{{\alpha_{0y} + \alpha_{1y} {h_y(k-1)}} + \beta_{1y} {\mathop{z_y^2}\nolimits}(k-1)}
\end{array}} \right],
\end{eqnarray}
%\begin{equation}
%\label{moh} h_y(k) = {\alpha_{0y} + \alpha_{1y} {h_y(k-1)}} + \beta_{1y} {\mathop{\rm q_y^2}\nolimits}(k-1)
%\end{equation}
where $\{\alpha_{0x}, \alpha_{1x}, \beta_{1x}\}$ and
$\{ \alpha_{0y}, \alpha_{1y}, \beta_{1y} \}$ are GARCH parameters of direction
$x$ and $y$, respectively, whereas $\alpha_0 = \theta \omega ({T}/{M})$,
$\alpha_1 = 1 - \theta ({T}/{M})$,
and $\alpha_2 = \xi \sqrt {{T}/{M}}$. According to (\ref{model1}), $z_x(k)$ and $z_y(k)$ describe as
$z_x(k) = \sqrt{h_x(k)} \varepsilon (k)$ and $z_y(k)=\sqrt{h_y(k)} \varepsilon (k)$
in direction $x$ and $y$, respectively, and $\varepsilon (k)$
is a sequence of zero-mean iid random variables with unity variance.

With regard to what has been described, (\ref{model19}) and
(\ref{eq_H2}) are reformed by augmenting $\mbox{\boldmath $h$}(k)$
to the state modeling of (\ref{model19}) in the form of the standard
Bayesian model \cite{hajiramezanali2018bayesian}. By defining an augmented state via some
manipulations, we can write:
\begin{equation}
\label{model30} \mbox{\boldmath $X$}_{N}(k)  = \mbox{\boldmath $F$}_{N} (\mbox{\boldmath $X$}(k-1) ,\mbox{\boldmath $h$}(k-1) ,T,\mu )
\end{equation}
where
\begin{eqnarray}
\nonumber
\mbox{\boldmath $X$}_{N}(k) &=& \left[ {\begin{array}{*{20}{c}}
   {{\mbox{\boldmath $X$}(k)}}  \\
   {{\mbox{\boldmath $h$}(k)}}  \\
\end{array}} \right], \\ \nonumber
\label{FN} \mbox{\boldmath $F$}_{N}  &=& \left[ {\begin{array}{*{20}{c}}
   {{\mbox{\boldmath $\Phi$}
   }(T,\mu){\mbox{\boldmath $X$}(k-1)} + {\mbox{\boldmath $U$}(k-1)}}  \\
   {h_x(k-1)}  \\
   {h_y(k-1)}  \\
\end{array}} \right],
\end{eqnarray}
which allows estimating both the target's kinematic parameters and its
stochastic volatility at the same time.

It is clear that timing is a significant factor in a real-time target tracking problem,
and many traditional algorithms demand to
target maneuvering detection as quickly as possible. But,
in the proposed GARCH model in (\ref{model30}),
we estimate the original state and stochastic volatility with a
filter simultaneously and unlike the decision based algorithms
such as IE, the maneuver detection procedure is eliminated.
Therefore, the consumption time of the maneuver detection is
zero. Now, we have a single non-linear and non-Gaussian model in
(\ref{model30}) whose solution via Kalman filtering results in
degraded performance. To resolve this difficulty, various
techniques have been investigated and applied in the literature
\cite{PF}, especially in the target tracking community. In the
next section, we suggest the bootstrap filtering to enhance the tracking accuracy
for the proposed non-liner non-Gaussian GARCH
maneuvering target model.

\section{Bootstrap Filter for Maneuvering Target Tracking}

Bayesian bootstrap filtering is an ideal approach suitable for
the simulation methodology. It provides a numerical solution to
calculate the distribution by a random sample vector. Therefore,
this feature makes it ideal where very complex densities can be
generated, like in the multiple model problems. Bootstrapping is
an iteration based algorithm  to obtain statistical properties of
a probability density function when only the samples from that density
are available. The bootstrap filter in \cite{bootstrap} performs
Bayesian estimation by predicting and updating a set of samples
representing the probability density functions of the system. The
samples tend to concentrate in high probability regions and the
evolution of the samples is computationally simple. Therefore,
this technique is very significant for recursive Bayesian
estimation of non-linear and non-Gaussian systems.

In bootstrapping, a probability
density function, $p(x)$, is approximated by a set of samples.
The bootstrap filter assumes that the
statistical properties of the samples are approximately
those of the density function by using the duality that exists between a sample and
the density that generated it.

\begin{figure}
\psset{unit=.5}
%\begin{center}
\centering
\begin{pspicture}(-9,-18)(8,5.5)
\psellipse[linewidth=.7pt](-2,4.5)(2,1)
\rput(-2,4.5){\small Initialization}
\psline[linewidth=.7pt]{->}(-2,3.5)(-2,2)
\psframe[linewidth=.7pt](-7.75,-1.25)(3.75,2)
\rput(-5.25,1.5){\small Time Propagation}
\rput(-2,0){\begin{small}$ \left[ {\begin{array}{*{20}{c}}
{\mbox{\boldmath $X$}^{(i)}_{k|k-1}}\\
{\mbox{\boldmath $h$}^{(i)}_{k|k-1}}
\end{array}} \right] = \mbox{\boldmath $F$}_N(\mbox{\boldmath $X$}^{(i)}_{k-1|k-1},\mbox{\boldmath $h$}^{(i)}_{k-1|k-1})$\end{small}}
\psline[linewidth=.7pt]{->}(-2,-1.25)(-2,-3.5)
\psline[linewidth=.7pt]{-}(-9,-1.6)(-2,-1.6)
\psline[linewidth=.7pt]{-}(-9,-1.6)(-9,-17)
\psline[linewidth=.7pt]{->}(-9,-17)(-5,-17)
\psframe[linestyle=dashed,linewidth=.9pt](-8.5,-2)(4.5,-10)
\rput(-5.75,-2.5){\small Weight Calculation}
\rput(-1.25,-9.5){\begin{small}$\omega^{(i)}_k$\end{small}}
\psframe[linewidth=.7pt](-7,-6.5)(3,-3.5)
\rput(-5,-4){\small Weight Update}
\rput(-2,-5.25){\begin{small}$\bar{\omega}^{(i)}_k = p\left( \mbox{\boldmath $y$}_k | \left[ {\begin{array}{*{20}{c}}
{\mbox{\boldmath $X$}^{(i)}_{k|k-1}}\\
{\mbox{\boldmath $h$}^{(i)}_{k|k-1}}
\end{array}} \right] \right)$\end{small}}
\psline[linewidth=.7pt]{->}(-2,-6.5)(-2,-7.5)
\psframe[linewidth=.7pt](-7,-9)(3,-7.5)
\rput(-2,-8.25){\small Weight Normalization}
\psline[linewidth=.7pt]{->}(-2,-9)(-2,-11)
\pspolygon[linewidth=.7pt](-5,-13)(-2,-15)(1,-13)(-2,-11)
\rput(-2,-13){\begin{small}$N_{eff}<0.5N_s$\end{small}}
\psline[linewidth=.7pt]{->}(-2,-15)(-2,-16)
\rput(-1.25,-15.5){\small Yes}
\psline[linewidth=.7pt]{->}(1,-13)(2,-13)
\rput(1.5,-12.5){\small No}
\psframe[linewidth=.7pt](-5,-18)(1,-16)
\rput(-2,-17){\small Bootstrap Resampling}
\psframe[linewidth=.7pt](2,-12)(8,-14)
\psline[linewidth=.7pt]{-}(5,-12)(5,2.75)
\psline[linewidth=.7pt]{->}(5,2.75)(-2,2.75)
\rput(5,-13){\small Output Estimation}
\psline[linewidth=.7pt]{-}(1,-17)(5,-17)
\psline[linewidth=.7pt]{->}(5,-17)(5,-14)
\rput(6.25,-11){\begin{small}$\left[ {\begin{array}{*{20}{c}}
{\mbox{\boldmath $\hat{X}$}_{k|k}}\\
{\mbox{\boldmath $\hat{h}$}_{k|k}}
\end{array}} \right]$\end{small}}
\end{pspicture}
\caption{\label{fig1}Functional structure of proposed PF-AR-GARCH method.}
%\end{center}
\label{fig}
\end{figure}
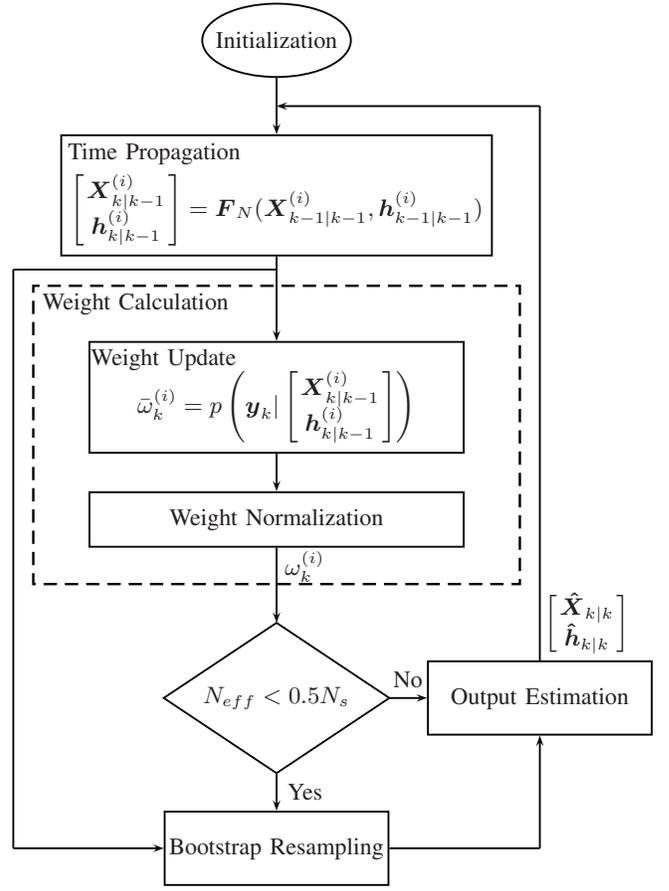
The proposed non-linear and non-Gaussian single model is
\begin{equation}
\label{boot} \mbox{\boldmath $X$}(k) = \mbox{\boldmath
$F$}_N(\mbox{\boldmath $X$}(k-1),\mbox{\boldmath$h$}(k), T, \mu),
\end{equation}
and the measurement equation can be written as
\begin{equation}
\label{bootmeas} \mbox{\boldmath $Y$}(k) = \mbox{\boldmath $\mathcal{H}$}\mbox{\boldmath $X$}(k) + \mbox{\boldmath$\epsilon$}(k),
\end{equation}
where $\mbox{\boldmath $Y$}(k) = [x(t)\quad y(t)\quad v_x(k)\quad
v_y(k)]^T$ and $\mbox{\boldmath$\epsilon$} = [\epsilon_x \quad
\epsilon_y \quad \epsilon_{v_x} \quad \epsilon_{v_y}]^T \sim
{\bf{\mathcal{N}}}(0,{\mbox{\boldmath $R$}})$. Here,
$\mbox{\boldmath $R$}$ is a $4 \times 4$ diagonal measurement
covariance matrix in (\ref{bootmeas}) with the diagonal elements
equal to the variances of the positions $x$ and $y,$ and
velocities $v_x$ and $v_y$, that are
$\sigma^2_{Position}=E\{\epsilon^2_x\}=E\{\epsilon^2_y\}$ and
$\sigma^2_{Velocity}=E\{\epsilon^2_{v_x}\}=E\{\epsilon^2_{v_y}\}$,
respectively. Therefore, $\mbox{\boldmath $\mathcal{H}$}$ in
(\ref{bootmeas}) is considered as
\begin{equation}
\nonumber \mbox{\boldmath $\mathcal{H}$} = \left[
{\begin{array}{*{20}{c}}
   \mbox{\boldmath $I$}_{4\times 4} &\:\:  \mbox{\boldmath $O$}_{4\times 4} \\
   \mbox{\boldmath $O$}_{4\times 4} &\:\: \mbox{\boldmath $O$}_{4\times 4} \\
\end{array}} \right]
\end{equation}
for our 8-dimensional proposed state model in (\ref{boot})
where $\mbox{\boldmath $I$}$ indicates the identity matrix and $\mbox{\boldmath $O$}$
shows the zero matrix.

If a set of $N_s$ particles, number of particles is an index of
number of iterations in bootstrap PF, $\mbox{\boldmath\small
$X$}(k-1|k-1)$, is approximately distributed as the prior density,
$p(\mbox{\boldmath\small $X$}(k-1)|\mbox{\boldmath\small
$Y$}(k-1))$ and a further set of samples,
$\mbox{\boldmath$h$}(k)$, approximately distributed as the process
noise density, $p_{GARCH}(\mbox{\boldmath\small $h$}(k))$, then
the set of samples, $\mbox{\boldmath\small $X$}(k|k-1)$, are
subsequently distributed approximately as the precision density
$p(\mbox{\boldmath\small $X$}(k)|\mbox{\boldmath\small
$Y$}(k-1))$. In this approach, the
predicted density particles have been generated without the need to
apply the multidimensional integration of Bayes'
rule which is the main
computational disadvantage of grid-based Bayesian estimators \cite{bootstrap}. Thus,
we can apply much larger sample sizes in order to obtain greater
accuracy without excessive computation \cite{bootstrap}.

\section{Complexity analysis}
In this section, the complexity of the proposed method is
evaluated. The main aim of this paper is to propose a single maneuvering target model
with an acceptable performance instead of using  MM.
Accordingly, we proposed GARCH model while we provided arguments about the superiority of this single model.
This model can be used successfully as one of the models in numerous MM
configurations.
Therefore, the tracking algorithm is not the only main objective in this article. Since
the proposed non-linear and non-Gaussian model is filtered
with the PF algorithm, we provide a fair comparison for the computational complexity of
the PF-AR-GARCH with some subcategories of MM methodology
such as auxiliary (AUX)-MMPF and bootstrap-MMPF which apply
the same filtering scheme (i.e. PF algorithm).
On the other hand, the EKF and UKF estimates are not accurate
for the non-linear non-Gaussian MM problems. Naturally, PF can
be used in IMM for the target tracking \cite{bootsrap2} and visual tracking \cite{visual} applications.
In a future research, the proposed GARCH model
can be filtered by the algorithms with lower complexity such as EKF, UKF, PF-UKF and etc.
We did not appeal to these filtering approaches because the
literature for non-stationary unscented
Kalman filtering approaches requires the
development.
Thereupon, our proposed GARCH model can be comparable with the lower complex MM such as EKF-IMM and UKF-IMM and etc.

Let $d$ denotes the dimension of the state vector, and $N_s$ refers to the number of particles.
As a first order approximation, the complexity of the bootstrap PF algorithm is of $\mathcal{O}(N_s d^2)$,
while the Kalman filter is of $\mathcal{O}(3 d^3)$. However, it is possible to implement the same PF procedure for MM in
 $ \mathcal{O}(N_s \sum_{i=1}^M d^2_i)$  operations, where $M$ is the number of models and
$d_i$ denotes the dimension of the state vector of each model.
According to the section III, formula (\ref{model30}), the dimension of the proposed state
model is $8$. This indicates that in the best case of the MM, the complexity
is about $1.2$ times more than the PF-AR-GARCH method in an
application with $M=2$, $d_1 = d_2 = 6$ and the same $N_s$. The difference becomes more apparent
when $M$ increases. For example, considering $M=3$, $d_1=d_2=d_3=6$, the complexity of MM
becomes about $1.7$ times more than the complexity of the proposed method.

When PF is used in practice, we often wish
to minimize the number of particles in order to reduce the computational
complexity. In this study, we selected $N_s = 50$ for our bootstrap PF-AR-GARCH model
algorithm according to the several different simulations. Moreover, the implemented IMM
includes three different models, i.e., a constant velocity (CV) model,  a constant
acceleration (CA) model, and a constant turn (CT) model.
 Therefore, according to the complexity of the Kalman filter
mentioned above, the complexity of the IMM is of
$\mathcal{O}(3\sum_{i=1}^M d^3_i)$. We can indicate analytically
that the complexity of the PF-AR-GARCH scheme is about $1.64$
times more than the complexity of the applied IMM. The difference
becomes less when the dimension of the measurement vector
increases, in which the measurement update of Kalman filter
becomes more complex. In addition, we should consider that in most
practical applications, the measurement equations are non-linear
and non-Gaussian. In non-linear and non-Gaussian problems, the
Kalman filter is not useful. Hence, the complexity of the IMM will
be increased in such situations and it will be comparable with the
proposed method. The MIE algorithm has lower complexity, however,
its performance is lower than the other approaches as shown in the
simulation results.

In Table \ref{Tab.complex}, we provide the number of operations for the PF-AR-GARCH model and
MM with two different resampling schemes, Bootstrap and auxiliary particle
filter. Although a precise comparison is difficult to make, we could conclude
that the proposed method has an acceptable complexity in comparison with MMs.
As it was mentioned above, the complexity of the proposed GARCH model can be reduced
if we apply the tracking algorithms
that possess the lower computational complexity for our proposed non-stationary model.
Several different methods such as UKF-GARCH model can be proposed in the future literatures
in order to reduce the computational complexity.
\begin{table}[t]
\caption{The number of operations for different algorithms} \label{Tab.complex}
\begin{center}
\begin{tabular}{|c|c|c|c|}
\cline {1-4}
%\multicolumn {1} {c|}{}
 & & &  \\
\multirow {-2}{2.2cm} {\centering Operations} & \multirow {-2}{2.1cm} {\centering Multiplications} & \multirow {-2}{1.3cm} {\centering Additions} & \multirow {-2}{1.3cm} {\centering Comparisons} \\ \hline
 & & &  \\
\multirow {-2}{2.4cm} {\centering Bootstrap PF-AR-GARCH}& \multirow {-2}{1.3cm} {\centering $(d^2 + 9)N_s$} &  \multirow {-2}{1.3cm} {\centering $17 N_s$}& \multirow {-2}{1.3cm} {\centering $2 N_s$} \\ \hline
 & & &  \\
\multirow {-2}{2.2cm} {\centering AUX PF-AR-GARCH}& \multirow {-2}{1.3cm} {\centering $\!\!\!\small{(d^2+15)N_s}$} &  \multirow {-2}{1.3cm} {\centering $20 N_s$}& \multirow {-2}{1.3cm} {\centering $3 N_s$} \\ \hline
 & & &  \\
\multirow {-2}{2.2cm} {\centering Bootstrap MMPF}& \multirow {-2}{1.3cm} {\centering $\!\!\!\!\!\!\!\!\!(\small{\sum_{i = 1}^M {d_i^2} + 9)N_s}$} &  \multirow {-2}{1.3cm} {\centering $17 M N_s$}& \multirow {-2}{1.3cm} {\centering $2 M N_s$} \\ \hline
 & & &  \\
\multirow {-2}{2.2cm} {\centering AUX MMPF}& \multirow {-2}{1.3cm} {\centering $\!\!\!\!\!\!\!\!\!\!(\small{\sum_{i = 1}^M {d_i^2} + 15)N_s}$} &  \multirow {-2}{1.3cm} {\centering $20 M N_s$}& \multirow {-2}{1.3cm} {\centering $3 M N_s$} \\ \hline
\end{tabular}
\end{center}
\end{table}

\section{simulation results}
In the following, some scenarios are  simulated to evaluate the proposed GARCH
model in comparison with several popular maneuvering target tracking algorithms.
These simulated target trajectories contain a range of maneuver and evasive target trajectories designed to
cover a wide range of target types and maneuver capabilities.
The performance of the new tracking scheme (PF-AR-GARCH technique) is compared with
the PF algorithm, the MIE method and the IMM algorithms.
The IMM algorithms consist of one CV
and one CA
models, on the other hand, EKF-IMM and UKF-IMM approaches consist of one CV, one CA, and one CT models.

\subsection{Test trajectory 1}
As an evaluation of the new tracking scheme
(PF-AR-GARCH method), two
examples cited in \cite{VD,IE,MIE,2}, will be
simulated with some slight changes to accommodate two dimension, in $X$ and $Y$ directions. Afterwards, we simulate
different examples drawn from some published results.

\subsubsection {Example A}
The position and velocity of the target are
measured every $T = 0.05 s$ and the process
noise is assumed to be zero for the trajectory simulation.
The initial position of the target is given by
$\boldmath{X}(t) = [2000 m,\, 10000 m,\, 0 m/s,\, -15 m/s].$
The variances of the
measurement noises are $\sigma_{Position}^2 = 10^4 m,$ and $\sigma_{Velocity}^2 = 25 m/s$.
The target is assumed to move with a constant velocity
until time $t = 5 s$ in $X$ direction, and the $Y$ direction of the velocity is constant until
time $ t = 3.5 s$. The target begins to
maneuver at an acceleration of $U_y = 38 m/s^2$ for the sample interval $[70, 200]$.
At the sample interval
$[100, 200]$, a constant acceleration $U_x = 40 m/s^2$
is applied in $X$ direction.
The process noise variances of IMM models, $\sigma_{CV}^2$ and $\sigma_{CA}^2$,
are selected according to \cite{IMM2001}.
In this
simulation, the standard deviation parameters are selected to be $\sigma_{CA} = 0.5 U_x = 20$ and $\sigma_{CA} = 0.1 \sigma_{CA} = 2,$ for the CA and
CV models, respectively.
The transition probability matrix
between the two models is given by $p_{ij} = [0.99\quad 0.01;\quad 0.01\quad 0.99],$ and $p_{ij} = [0.85\quad 0.15;\quad 0.15\quad 0.85]$ for
IMM-1 and IMM-2, respectively.
According to the best scenario, the variance parameters of the PF scheme
and the MIE method are selected to be $\sigma_{PF}^2 = 150$ and $\sigma_{MIE}^2 = 100$, respectively.
The number of the particles related to the PF algorithm is selected to be $N_s = 200$.

\begin{figure}[t]
\center
\begin{minipage}[b]{1.0\linewidth}
\centering
\centerline{\includegraphics[scale=.4]{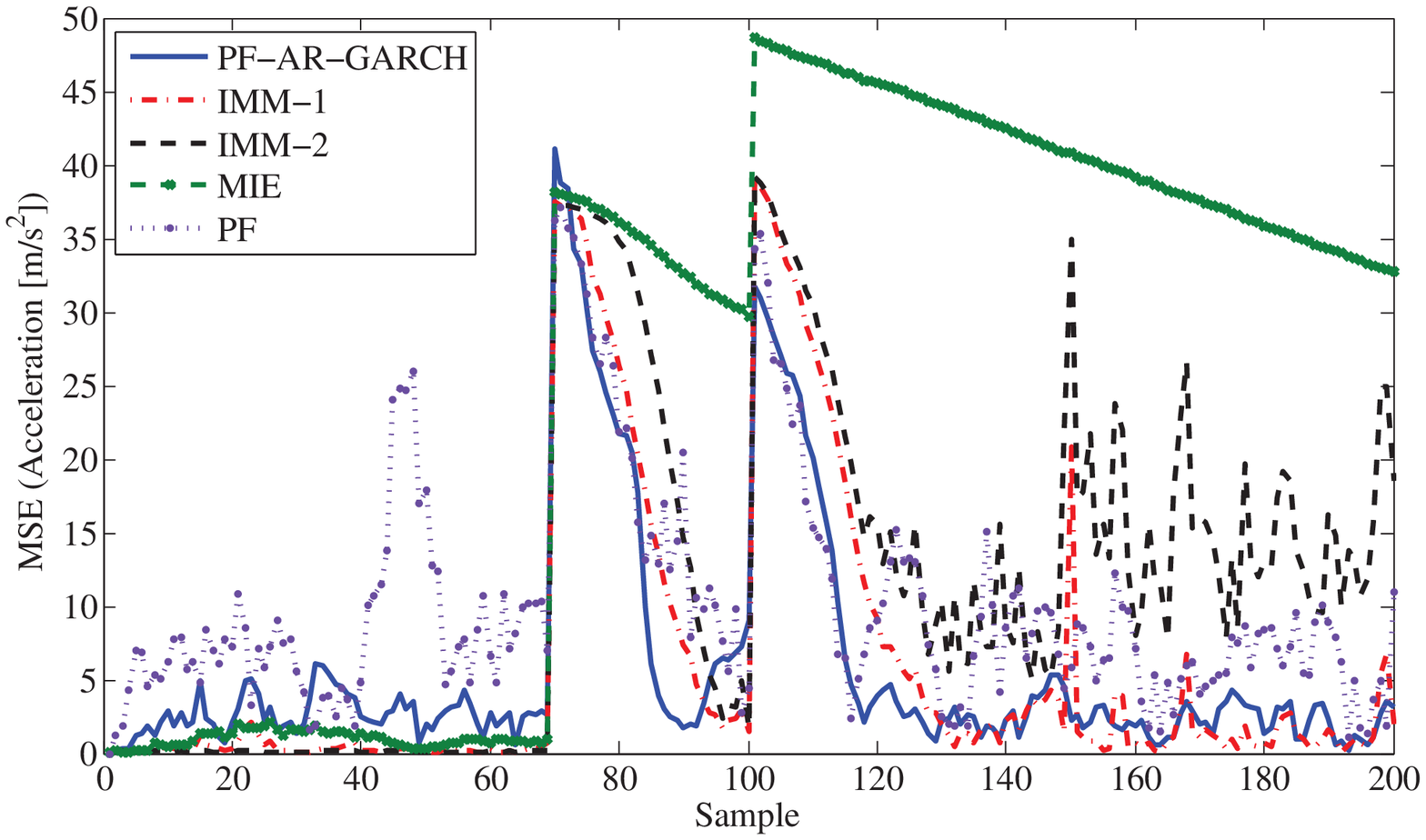}}
\centerline{(a)}
\end{minipage}
\hfil
\begin{minipage}[b]{1.0\linewidth}
\centering
\centerline{\includegraphics[scale=.4]{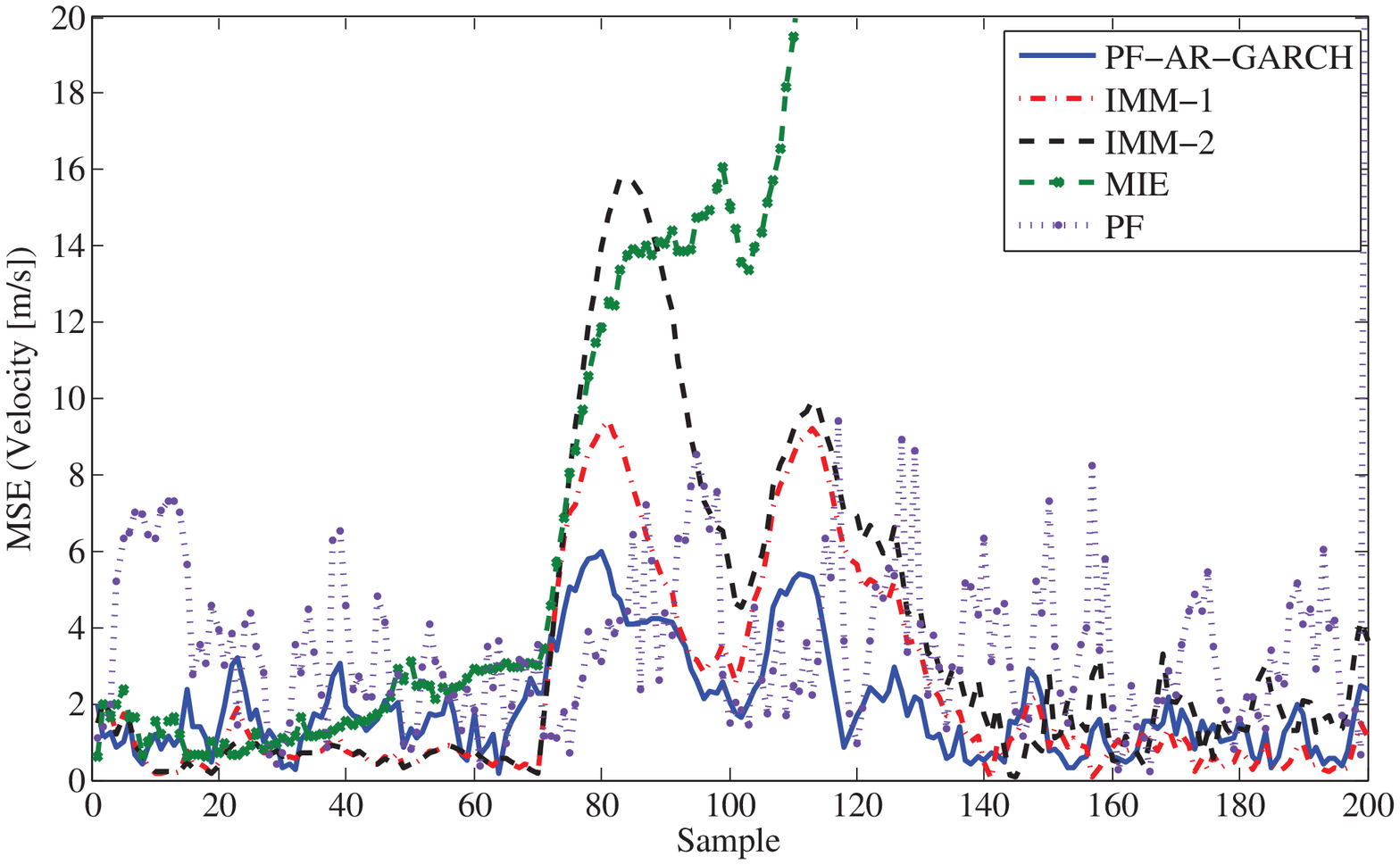}}
\centerline{(b)}
\end{minipage}
\caption{The MSE of different methods for (a) acceleration and (b) velocity}
\label{IEMSE}
\end{figure}

The performance of the IMM algorithms, MIE scheme, PF approach and
the PF-AR-GARCH method are shown in Fig. \ref{IEMSE} for this simulation.
It can be seen that the PF-AR-GARCH
method outperforms the IMM schemes
when the target is in the transition period.
In the steady state of the accelerating
interval, some advantages can also be obtained through the proposed PF-AR-GARCH model but the improvement is generally
not so significant as that in the transition case.
In addition, the effect of transition probability matrix is obvious, especially in the constant acceleration period
for the acceleration estimation and in the transition interval for the velocity estimation.
Although the performance of the IMM methods is better than the proposed method in the CV interval,
the proposed PF-AR-GARCH method outperforms in the other time intervals, i.e. the transition periods and CA intervals, which are more
important in the maneuver target tracking applications.

In comparison with MIE model, the PF-AR-GARCH algorithm has a better
performance in transition and constant acceleration periods.
In Fig. \ref{IEMSE}(a), the performance of MIE is eliminated after 100
samples due to an unacceptable performance and make the figure more clear.
The PF algorithm shows weak performance in the CV interval compared
with the other approaches. Although the performance of the
PF is better than IMMs and MIE algorithms in the transition period especially for velocity estimation,
the performance of PF-AR-GARCH approach is superior than PF
for acceleration and velocity estimation in this period. The estimation delay for the PF is illustrated
in the transition period. However, the PF performance is degraded in the CA interval
in comparison with the proposed PF-AR-GARCH method.

\begin{table}[t]
\caption{The RMSE Result of Different Methods by 1000 Monte Carlo Simulations for Test Trajectory 1 (High Maneuver)} \label{Tab.CVCA}
\begin{center}
\begin{tabular}{|c|c|c|c|}
\cline {1-4}
%\multicolumn {1} {c|}{}
  & & &  \\
\multirow {-2}{1.3cm} {\centering RMSE} & \multirow {-2}{1.3cm} {\centering Position [$m$]} & \multirow {-2}{1.3cm} {\centering Velocity [$m/s$]} & \multirow {-2}{1.3cm} {\centering Acceleration [$m/s^2$]} \\ \hline
 & & &  \\
\multirow {-2}{2.2cm} {\centering PF-AR-GARCH}& \multirow {-2}{1.3cm} {\centering 2.0566} &  \multirow {-2}{1.3cm} {\centering 1.2849}& \multirow {-2}{1.3cm} {\centering 2.4288} \\ \hline
 & & &  \\
\multirow {-2}{2.2cm} {\centering IMM-1}& \multirow {-2}{1.3cm} {\centering 2.0732} &  \multirow {-2}{1.3cm} {\centering 1.5256}& \multirow {-2}{1.3cm} {\centering 2.5194} \\ \hline
 & & &  \\
\multirow {-2}{2.2cm} {\centering IMM-2}& \multirow {-2}{1.3cm} {\centering 2.2313} &  \multirow {-2}{1.3cm} {\centering 1.8617}& \multirow {-2}{1.3cm} {\centering 3.4251} \\ \hline
 & & &  \\
\multirow {-2}{2.2cm} {\centering MIE}& \multirow {-2}{1.3cm} {\centering 5.4701} &  \multirow {-2}{1.3cm} {\centering 5.9399}& \multirow {-2}{1.3cm} {\centering 5.0973} \\ \hline
 & & &  \\
\multirow {-2}{2.2cm} {\centering PF}& \multirow {-2}{1.3cm} {\centering 4.6599} &  \multirow {-2}{1.3cm} {\centering 1.9908}& \multirow {-2}{1.3cm} {\centering 3.3066} \\ \hline
\end{tabular}
\end{center}
\end{table}
\begin{figure}[t]
\center
\centerline{\includegraphics[scale=.4]{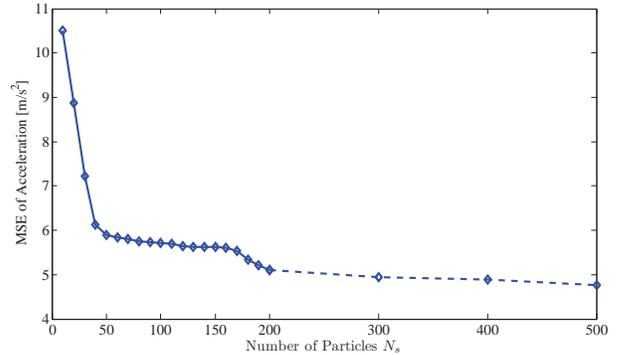}}
\caption{The Number of Particles versus the MSE of PF-AR-GARCH method for Test Trajectory 1 (High Maneuver)}
\label{MSENH}
\end{figure}

The root mean square error (RMSE) of the proposed method in comparison with the
IMM-1, IMM-2, MIE and PF algorithms, is shown in Table \ref{Tab.CVCA}.
Some Monte Carlo simulations with $1000$ runs in each simulation are
performed.
It is seen that the improvement, due
to the proposed PF-AR-GARCH, is rather significant, especially in
the velocity and acceleration estimations.
As we mentioned in the introduction, more accurate estimations
in the velocity and acceleration
are particularly useful in some applications
such as threat evaluation, etc.
According to the PF results in Fig. \ref{IEMSE} and Table \ref{Tab.CVCA}, we can conclude that
the PF algorithms without considering the proper model does not have
good performance. This test illustrates the significant effect of
the proposed GARCH model for the maneuvering target tracking applications.

Fig. \ref{MSENH} illustrates the effect of number of particles for the bootstrapped PF-AR-GARCH tracking
in the test trajectory $1$ when the other parameters are fixed. The more
$N_s$ provide the less RMSE for the proposed method
while the IMM and MIE are not dependent on this degree of freedom, but,
this higher accuracy provided by bootstrapping comes at an extra
computational cost as denoted in Table I.
According to the Fig. \ref{MSENH},
$N_s = 50$ is selected to reduce the computational complexity in
this simulation.

\subsubsection {Example B}
\begin{figure}[t]
\center
\centerline{\includegraphics[scale=.4]{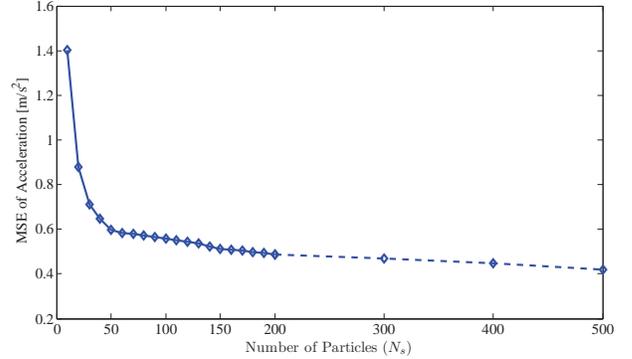}}
\caption{The Number of Particles versus the MSE of PF-AR-GARCH method for Test Trajectory 1 (Low Maneuver)}
\label{MSENL}
\end{figure}

In this example, we consider the sampling time is
$T = 0.05 s$, the initial position of target is $\boldmath{X}(t) = [2000 m,\, 10000 m,\, 0 m/s,\, -15 m/s],$
the variances of the
measurement noises are $\sigma_{Position}^2 = 10^4 m,$ and $\sigma_{Velocity}^2 = 25 m/s$, and the process
noise is zero for the trajectory simulation.
The target is assumed to move with a constant velocity
until time $t = 5 s$ in $X$ direction, and the $Y$ direction of the velocity is constant until
time $ t = 3.5 s$. The target begins to
maneuver at an acceleration of $U_y = 1 m/s^2$ for the sample interval $[70, 200]$.
In the sample interval
$[100, 200]$, a constant acceleration $U_x = 0.8 m/s^2$
is applied in $X$ direction.
In this
simulation, $N_s = 50$ is selected based on Fig. \ref{MSENL}.
The standard deviation parameters of IMM are selected to be $\sigma_{CA} = 1 U_y = 1$ and $\sigma_{CV} = 0.1 \sigma_{CA} = 0.1,$ for the CA and
CV models, respectively.
The transition probability matrix
between the two models is given by $p_{ij} = [0.99\quad 0.01;\quad 0.01\quad 0.99].$
According to the best scenario, the variance parameters of the PF
and MIE methods are selected to be $\sigma_{PF}^2 = 5$ and $\sigma_{MIE}^2 = 5$, respectively.
The number of particles in PF algorithms is selected to be $N_s = 200$.

This example describes a low maneuver behavior to demonstrate the performance of
PF-AR-GARCH approach compared with the other methods.
Table \ref{Tab.CVCA2} shows the RMSE of proposed method in comparison with the IMM, MIE and PF algorithms.
Some Monte Carlo simulations with $1000$ runs in each simulation are
performed.
It is seen that the proposed PF-AR-GARCH scheme has a better performance
for an approximately non-jumpy situation.
Therefore, we conclude that PF-AR-GARCH is appropriate equivalently for low and high maneuver behaviors.

\begin{table}[t]
\caption{The RMSE Result of Different Methods by 1000 Monte Carlo Simulations for Test Trajectory 1 (Low Maneuver)} \label{Tab.CVCA2}
\begin{center}
\begin{tabular}{|c|c|c|c|}
\cline {1-4}
%\multicolumn {1} {c|}{}
  & & &  \\
\multirow {-2}{1.3cm} {\centering RMSE} & \multirow {-2}{1.3cm} {\centering Position [$m$]} & \multirow {-2}{1.3cm} {\centering Velocity [$m/s$]} & \multirow {-2}{1.3cm} {\centering Acceleration [$m/s^2$]} \\ \hline
 & & &  \\
\multirow {-2}{2.2cm} {\centering PF-AR-GARCH}& \multirow {-2}{1.3cm} {\centering 2.0462} &  \multirow {-2}{1.3cm} {\centering 1.1631}& \multirow {-2}{1.3cm} {\centering 0.7735} \\ \hline
 & & &  \\
\multirow {-2}{2.2cm} {\centering IMM}& \multirow {-2}{1.3cm} {\centering 2.1622} &  \multirow {-2}{1.3cm} {\centering 1.2897}& \multirow {-2}{1.3cm} {\centering 1.0211} \\ \hline
 & & &  \\
\multirow {-2}{2.2cm} {\centering MIE}& \multirow {-2}{1.3cm} {\centering 9.0189} &  \multirow {-2}{1.3cm} {\centering 3.7593}& \multirow {-2}{1.3cm} {\centering 1.6243} \\ \hline
 & & &  \\
\multirow {-2}{2.2cm} {\centering PF}& \multirow {-2}{1.3cm} {\centering 4.6938} &  \multirow {-2}{1.3cm} {\centering 1.7110}& \multirow {-2}{1.3cm} {\centering 1.4165} \\ \hline
\end{tabular}
\end{center}
\end{table}

\subsection{Test trajectory 2}
In this test, the trajectory is simulated by
heavy-tailed property assumption of the state equation noise.
As we noted in the introduction,
there are some rich literatures on modeling heavy-tailed
system noise \cite{levy,USA}.
Our main purpose is to develop a model where a
maneuver can be considered as independent heavy-tailed
noise to one or more of the state variables.
The reason for this assumption is that the unimodal
heavy-tail distribution represents
usually small fluctuations and abrupt changes of acceleration
in a simultaneous manner.

For this test problem, the target's motion equation is
expressed in terms of the non-Gaussian noise as
follows:
\begin{equation}
\nonumber \mbox{\boldmath $\dot X$}(t) =
\mbox{\boldmath $F$}\mbox{\boldmath $X$}(t) + \mbox{\boldmath $G$}\mbox{\boldmath $\vartheta$}(t)
\end{equation}
where $\mbox{\boldmath $X$}(t)=\left[\mbox{\boldmath $x$}(t)^T\:
\mbox{\boldmath $v$}(t)^T\: \mbox{\boldmath $a$}(t)\right]^T$
denotes a six-dimensional position-velocity-acceleration
parameter vector
and $\mbox{\boldmath $\vartheta$}(t)=[\vartheta_x(t)\:
\vartheta_y(t)]^T$
is the non-Gaussian system noise vector.
In this simulation trial, the true target trajectory is
simulated with independent random initial position,
velocity, and acceleration. The true process noise
is set to be student-$t$ with one degree of freedom.
The position and velocity of the target are measured
every $T = 1 s$. We have set the parameters value of
IMM as $\sigma_{CV}^2 = 0.1,$ $\sigma_{CA}^2 = 10$, and $p_{ij} = [0.8\quad 0.2;\quad 0.2\quad 0.8]$.
On the other hand, the parameters of the EKF-IMM and UKF-IMM approaches are selected to be
$\sigma_{CV}^2 = 1$, $\sigma_{CA}^2 = 5$ and $\sigma_{CT}^2 = 10$.
In addition, the transition probability matrix
between the three models in the EKF-IMM and UKF-IMM methods is given by
$p_{ij} = [0.99\quad 0.01\quad 0.00; 0.33\quad 0.34\quad 0.33; 0.00\quad 0.01\quad 0.99]$.
In this simulation, the selection of IMM parameters
is based on a trade off between the performance of steady state
and transition estimations as well as the previous test.
The MIE parameters are selected according to the best
performance as $\sigma^2 = 100$.

\begin{table}[t]
\caption{The RMSE Result of Different Methods by 1000 Monte Carlo Simulations for  Test Trajectory 2 (Student-$t$ Distribution)} \label{Tab.Cauchy}
\begin{center}
\begin{tabular}{|c|c|c|c|}
\cline {1-4}
%\multicolumn {1} {c|}{}
  & & &  \\
\multirow {-2}{1.3cm} {\centering RMSE} & \multirow {-2}{1.3cm} {\centering Position [$m$]} & \multirow {-2}{1.3cm} {\centering Velocity [$m/s$]} & \multirow {-2}{1.3cm} {\centering Acceleration [$m/s^2$]} \\ \hline
 & & &  \\
\multirow {-2}{2.2cm} {\centering PF-AR-GARCH}& \multirow {-2}{1.3cm} {\centering 6.0612} &  \multirow {-2}{1.3cm} {\centering 2.0857}& \multirow {-2}{1.3cm} {\centering 1.5900} \\ \hline
 & & &  \\
\multirow {-2}{2.2cm} {\centering IMM}& \multirow {-2}{1.3cm} {\centering 6.4961} &  \multirow {-2}{1.3cm} {\centering 2.5270}& \multirow {-2}{1.3cm} {\centering 2.4861} \\ \hline
 & & &  \\
\multirow {-2}{2.2cm} {\centering EKF-IMM}& \multirow {-2}{1.3cm} {\centering 6.4458} &  \multirow {-2}{1.3cm} {\centering 2.5204}& \multirow {-2}{1.3cm} {\centering 2.3709} \\ \hline
 & & &  \\
\multirow {-2}{2.2cm} {\centering UKF-IMM}& \multirow {-2}{1.3cm} {\centering 6.2996} &  \multirow {-2}{1.3cm} {\centering 2.5052}& \multirow {-2}{1.3cm} {\centering 2.1559} \\ \hline
 & & &  \\
\multirow {-2}{2.2cm} {\centering MIE}& \multirow {-2}{1.3cm} {\centering 41.7349} &  \multirow {-2}{1.3cm} {\centering 9.2089}& \multirow {-2}{1.3cm} {\centering 3.0882} \\ \hline
 & & &  \\
\multirow {-2}{2.2cm} {\centering PF}& \multirow {-2}{1.3cm} {\centering 30.8065} &  \multirow {-2}{1.3cm} {\centering 7.5206}& \multirow {-2}{1.3cm} {\centering 2.7412} \\ \hline
\end{tabular}
\end{center}
\end{table}
\begin{figure}[t]
\center
\centerline{\includegraphics[scale=.4]{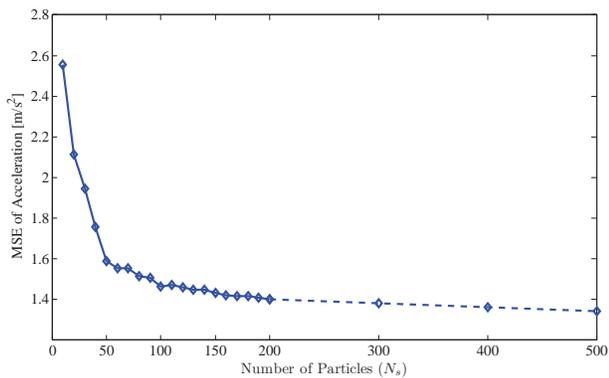}}
\caption{The Number of Particles versus the MSE of PF-AR-GARCH method for Test Trajectory 2}
\label{MSEN2}
\end{figure}

For the quantitative comparison,
the RMSE of tracking results are calculated. To achieve more
precise results,
some Monte Carlo simulations with 1000 runs is obtained
in each algorithm
and the RMSE values of the estimation
are computed by averaging. The results are illustrated in Table \ref{Tab.Cauchy}.
We can draw two important conclusions from Table \ref{Tab.Cauchy}.
First, the proposed GARCH approach can handle non-Gaussianity more properly.
Second, the effect of estimation delay in the MIE for high maneuvering trajectories causes large RMSE.

Fig. \ref{MSEN2} illustrates the effect of particles number for the bootstrapped PF-AR-GARCH tracking
in test trajectory $2$ when the other parameters are fixed.
According to the Fig. \ref{MSEN2},
the number of particles, $N_s$ is selected $50$ in order to reduce the computational complexity in
this test trajectory.

\subsection{Test trajectory 3}
The main goal of this study is (a) to show that the proposed model is so general that
can describe several different motion models, and (b)
to illustrate the sensitivity of IMM method
to the priori parameters, i.e. variance of the state models and the model transition probability matrix.
Moreover, this simulation shows that the dynamics are difficult to break up
into several different motion models in IMM methods. When the parameters
are changed to get better performance in one situation, the other
interval of the tracking have a bad performance and vice versa.
In order to provide better understanding of the IMM influence design variables,
three IMMs are simulated. We considered the noise standard deviation of the state models of IMM-1, IMM-2 and IMM-3
as $\sigma_{CV} = 0.1 \sigma_{CA}^2 = 1.5$, $\sigma_{CV} = 0.1 \sigma_{CA} = 0.8$, and $\sigma_{CV} = 0.1 \sigma_{CA} = 0.5$, respectively.
The model transition probability matrix $p_{ij} = [0.85\quad 0.15;\quad 0.15\quad 0.85]$ is considered for three IMMs.
The noise variance of the MIE method is considered as $\sigma_X^2 = 1000$.
The sampling time is $T = 1 s$ and the number of the
samples are 200. Initial position is
assumed to be $\mbox{\boldmath$X$}(0) = [10 m, -10 m, 10 m/s, 15 m/s]^T$.
$N_s = 400$ shows the best performance in the PF method for this scenario.

\begin{table}[t]
\caption{The RMSE Result of Different Methods by 1000 Monte Carlo simulations for Test Trajectory 3} \label{Tab.sin}
\begin{center}
\begin{tabular}{|c|c|c|c|}
\cline {1-4}
%\multicolumn {1} {c|}{}
  & & &  \\
\multirow {-2}{1.3cm} {\centering RMSE} & \multirow {-2}{1.3cm} {\centering Position [$m$]} & \multirow {-2}{1.3cm} {\centering Velocity [$m/s$]} & \multirow {-2}{1.3cm} {\centering Acceleration [$m/s^2$]} \\ \hline
 & & &  \\
\multirow {-2}{2.2cm} {\centering PF-AR-GARCH}& \multirow {-2}{1.3cm} {\centering 2.3029} &  \multirow {-2}{1.3cm} {\centering 1.6263}& \multirow {-2}{1.3cm} {\centering 1.9365} \\ \hline
 & & &  \\
\multirow {-2}{2.2cm} {\centering IMM-1}& \multirow {-2}{1.3cm} {\centering 3.7540} &  \multirow {-2}{1.3cm} {\centering 3.1086}& \multirow {-2}{1.3cm} {\centering 2.1046} \\ \hline
 & & &  \\
\multirow {-2}{2.2cm} {\centering IMM-2}& \multirow {-2}{1.3cm} {\centering 2.6401} &  \multirow {-2}{1.3cm} {\centering 2.4952}& \multirow {-2}{1.3cm} {\centering 2.1206} \\ \hline
 & & &  \\
\multirow {-2}{2.2cm} {\centering IMM-3}& \multirow {-2}{1.3cm} {\centering 3.2011} &  \multirow {-2}{1.3cm} {\centering 2.9341}& \multirow {-2}{1.3cm} {\centering 2.2144} \\ \hline
 & & &  \\
\multirow {-2}{2.2cm} {\centering MIE}& \multirow {-2}{1.3cm} {\centering 20.7740} &  \multirow {-2}{1.3cm} {\centering 3.9147}& \multirow {-2}{1.3cm} {\centering 3.3100} \\ \hline
 & & &  \\
\multirow {-2}{2.2cm} {\centering PF}& \multirow {-2}{1.3cm} {\centering 8.3287} &  \multirow {-2}{1.3cm} {\centering 3.7210}& \multirow {-2}{1.3cm} {\centering 2.9695} \\ \hline
\end{tabular}
\end{center}
\end{table}
\begin{figure}[t]
\center
\centerline{\includegraphics[scale=.4]{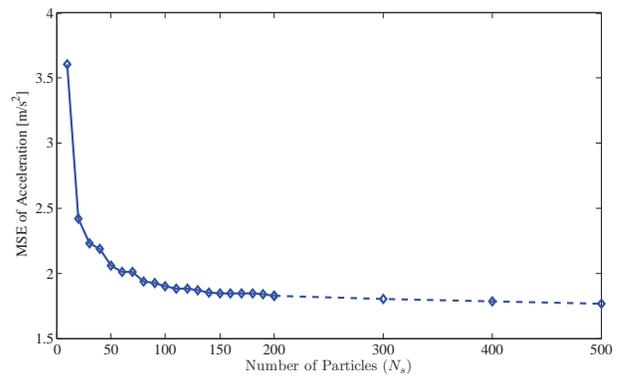}}
\caption{The Number of Particles versus the MSE of PF-AR-GARCH method for Test Trajectory 3}
\label{MSEN3}
\end{figure}

\begin{figure}[t]
\center
\begin{minipage}[b]{1.0\linewidth}
\centering
\centerline{\includegraphics[scale=.4]{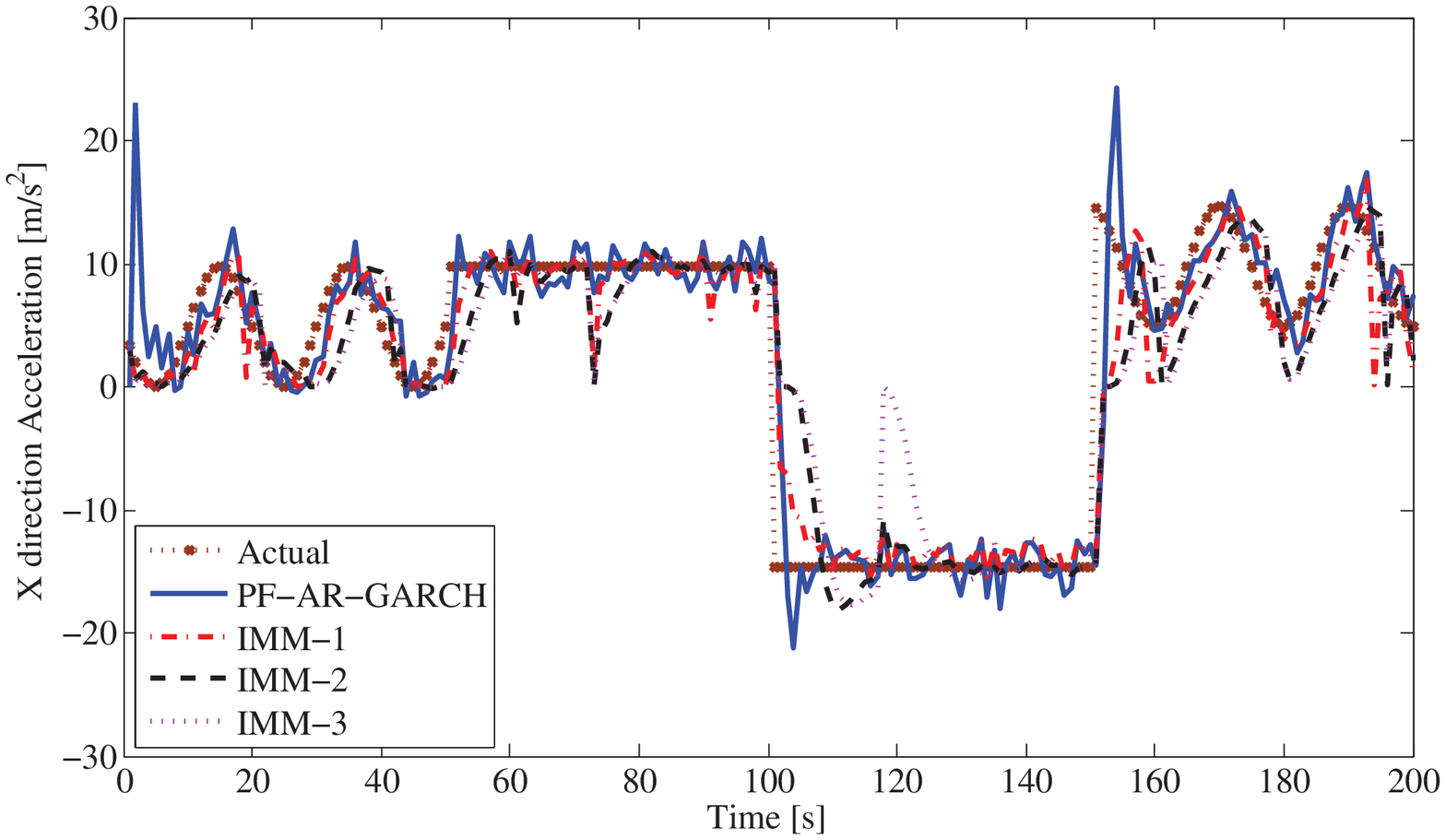}}
\centerline{(a)}
\end{minipage}
\hfil
\begin{minipage}[b]{1.0\linewidth}
\centering
\centerline{\includegraphics[scale=.4]{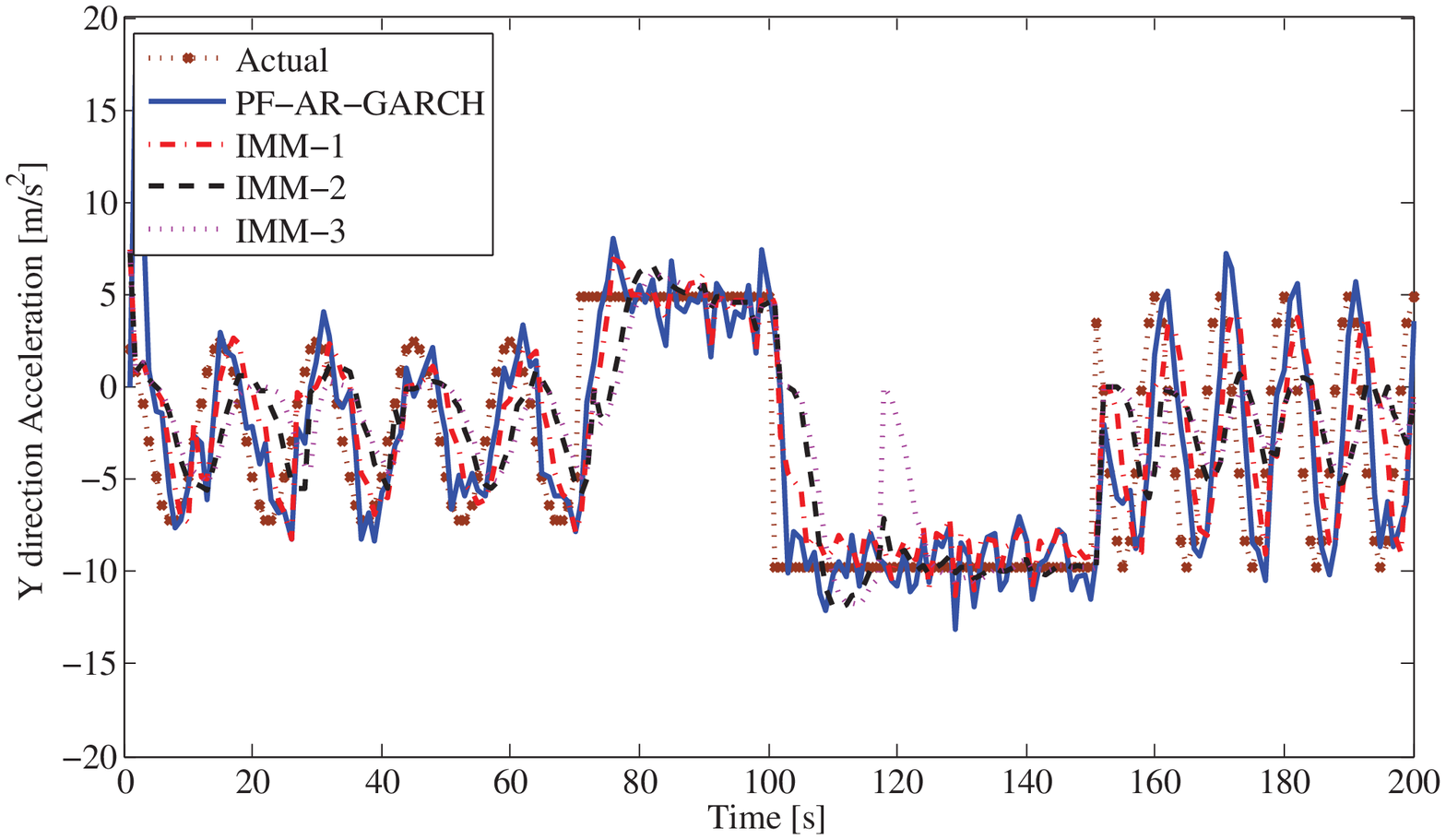}}
\centerline{(b)}
\end{minipage}
\caption{The actual and the estimated acceleration in (a) x direction and (b) y direction}
\label{sinacceleration}
\end{figure}

\begin{figure}[t]
\center
\begin{minipage}[b]{1.0\linewidth}
\centering
\centerline{\includegraphics[scale=.4]{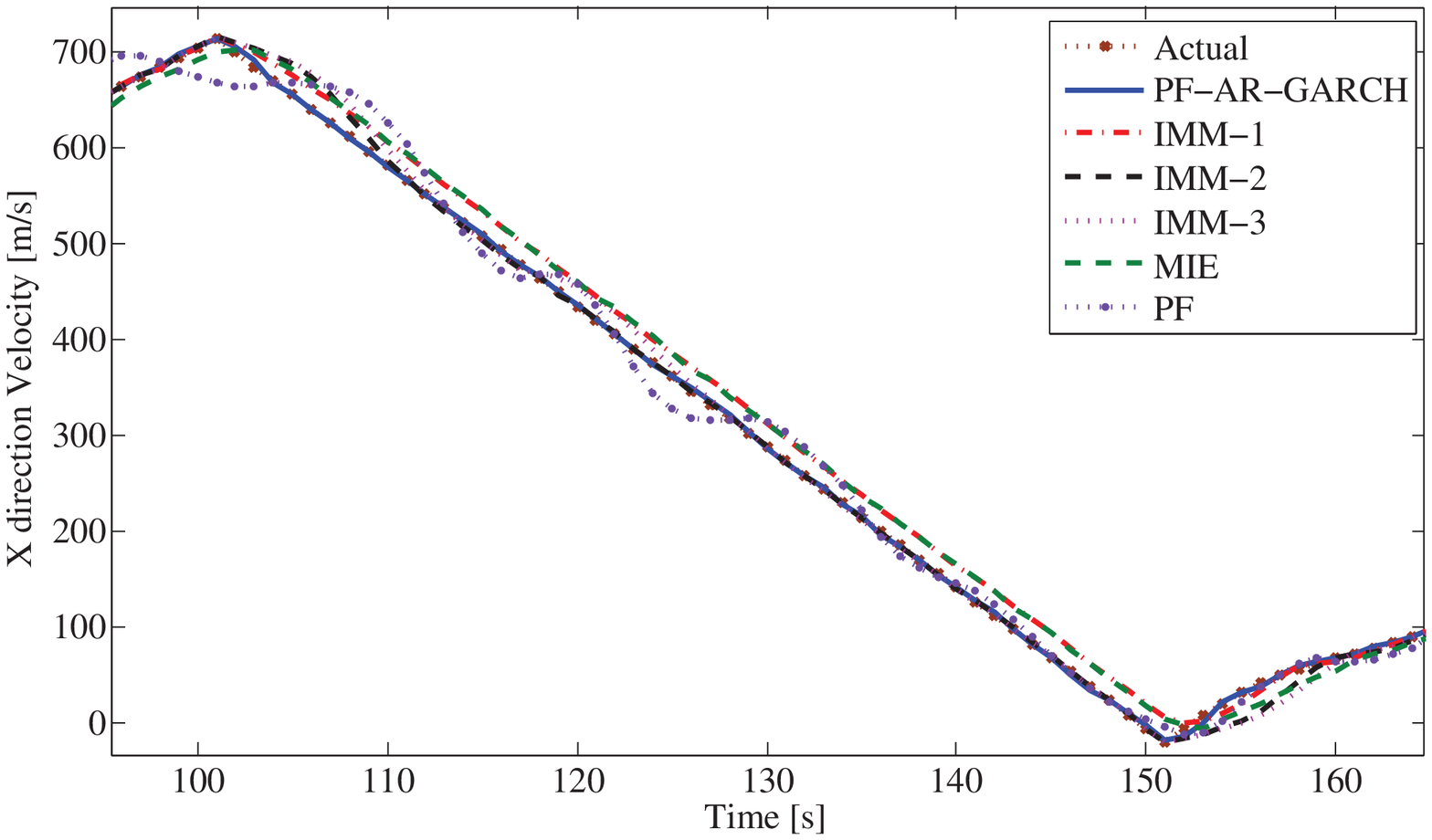}}
\centerline{(a)}
\end{minipage}
\hfil
\begin{minipage}[b]{1.0\linewidth}
\centering
\centerline{\includegraphics[scale=.4]{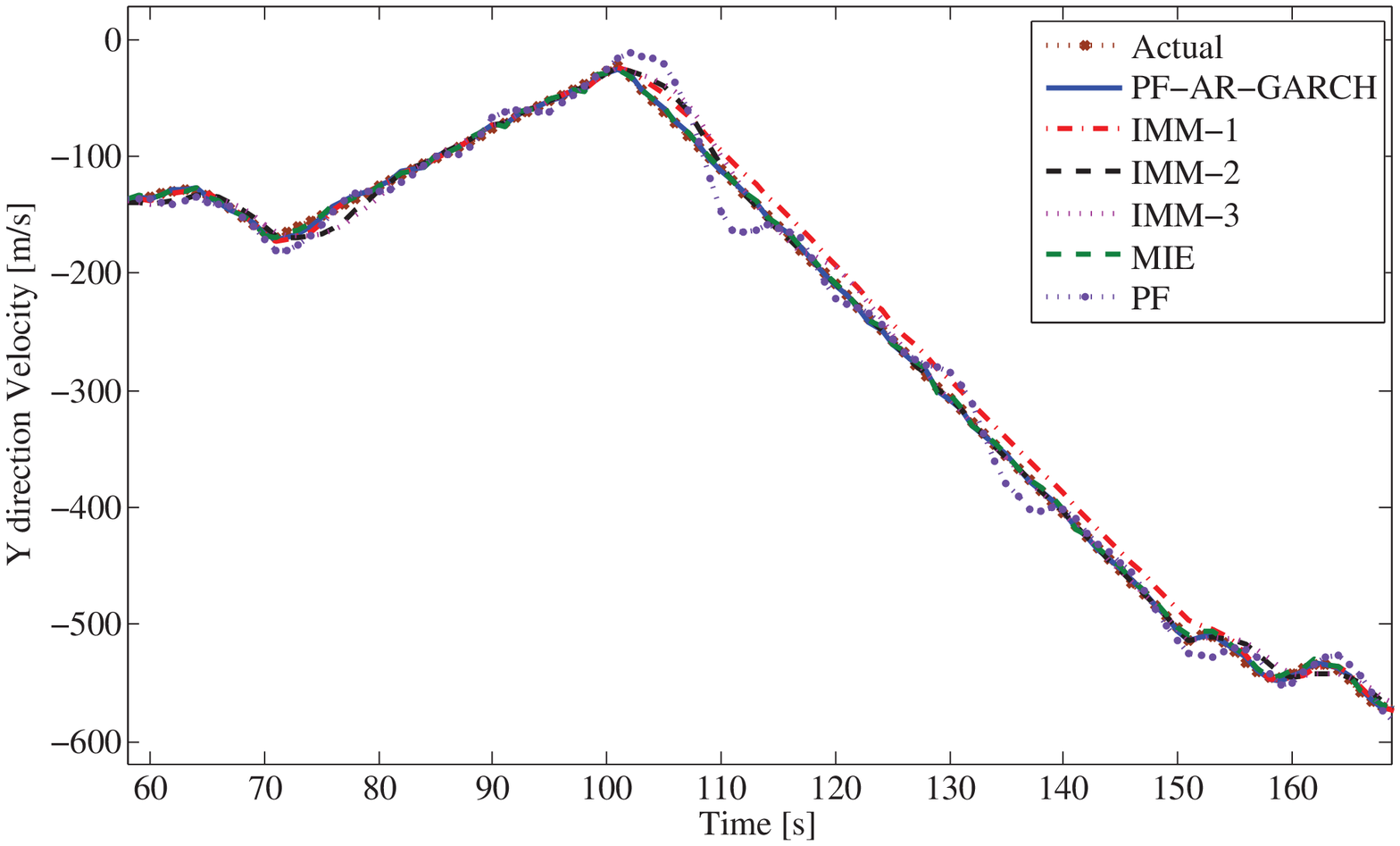}}
\centerline{(b)}
\end{minipage}
\caption{The actual and the estimated velocity in (a) x direction and (b) y direction}
\label{sinvelocity}
\end{figure}

Fig. \ref{MSEN3} illustrates the effect of number of particles for the bootstrapped PF-AR-GARCH tracking
in test trajectory $3$ when the other parameters are fixed.
According to the Fig. \ref{MSEN3}, $N_s = 100$ is selected in order to reduce the computational complexity in
this test trajectory.
Fig. \ref{sinacceleration} and Fig. \ref{sinvelocity} show the estimated and actual values
for acceleration and velocity of the target, respectively, which
are achieved by the PF-AR-GARCH, IMM, MIE and PF approaches.
Moreover, each algorithm is simulated 1000 times and the final
RMSE is illustrated in Table \ref{Tab.sin}.
Since the performances of the MIE and PF algorithms are not acceptable, their results are eliminated from Fig. \ref{sinacceleration}
to improve the quality of this figure.
The simulation
results show that
the proposed algorithm outperforms IMM methods, especially in sinusoidal case.
According to the Fig. \ref{sinacceleration}, at the time $t = 100 s$, the target acceleration changes drastically.
The GARCH model follows this sinusoidal trajectory quite well, whereas IMM does not.
In the time interval $100 s < t < 150 s$ after the abrupt changes in time instant $t = 100 s$, the value of conditioned likelihood
function in the CV model is extremely larger than the CA model for IMM-1. Then, the IMM method allocates higher
weights to the CV model in this time interval, near one.
Therefore, the difference of velocity is used as the estimation of acceleration on this time slot.
This approach provides a better performance to acceleration estimation of IMM-1.
Table \ref{Tab.sin} demonstrates that the IMM-1 has a good performance in the acceleration estimation compared with
the other IMMs, however, its performance to the position and velocity estimations is lower than IMM-2 and IMM-3.
Fig. \ref{sinvelocity} shows this unacceptable performance of IMM-1 to estimate the position and velocity.
According to Fig. \ref{sinvelocity}, the velocity estimation of IMM-1 is weak.
These differences among three IMMs tracking accuracies provide a conclusion that
IMM algorithm is sensitive to the priori information.
The variance of process noise of the IMM-1
results in a weak performance to velocity estimation
in the interval $100 s < t < 150 s$, whereas, an acceptable accuracy
to acceleration estimation is resulted
in the sinusoidal intervals. Moreover, the performance of the IMM-2 is increased in the constant acceleration with
decrease in the variance of the process noise. However, its performance is degraded in the sinusoidal interval.
The decrease of the noise variance in the IMM-3 shows an inappropriate effect in the $100 s < t < 150 s$ interval.

\begin{figure}[t]
\center
\begin{minipage}[b]{1.0\linewidth}
\centering
\centerline{\includegraphics[scale=.4]{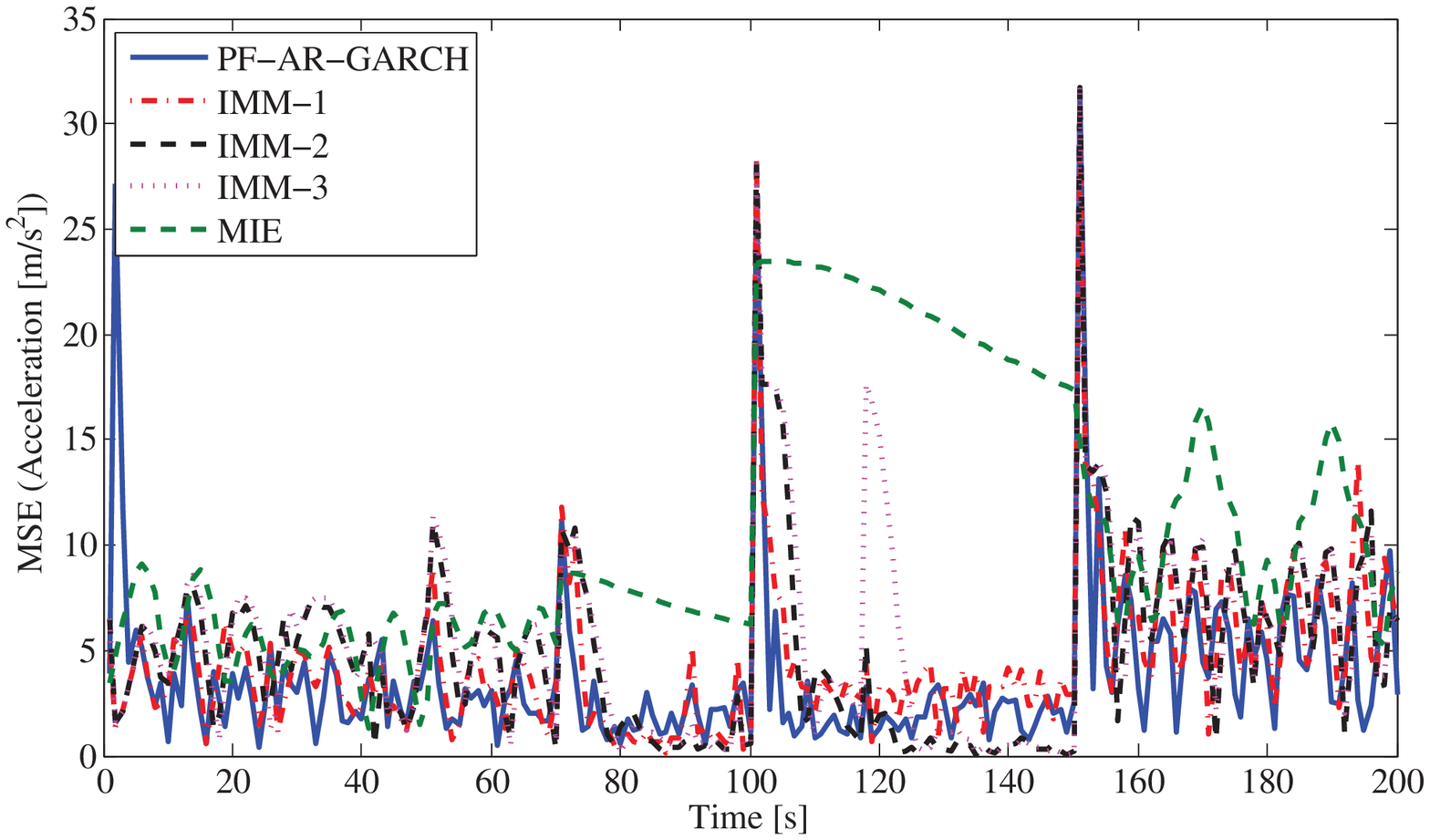}}
\centerline{(a)}
\end{minipage}
\hfil
\begin{minipage}[b]{1.0\linewidth}
\centering
\centerline{\includegraphics[scale=.4]{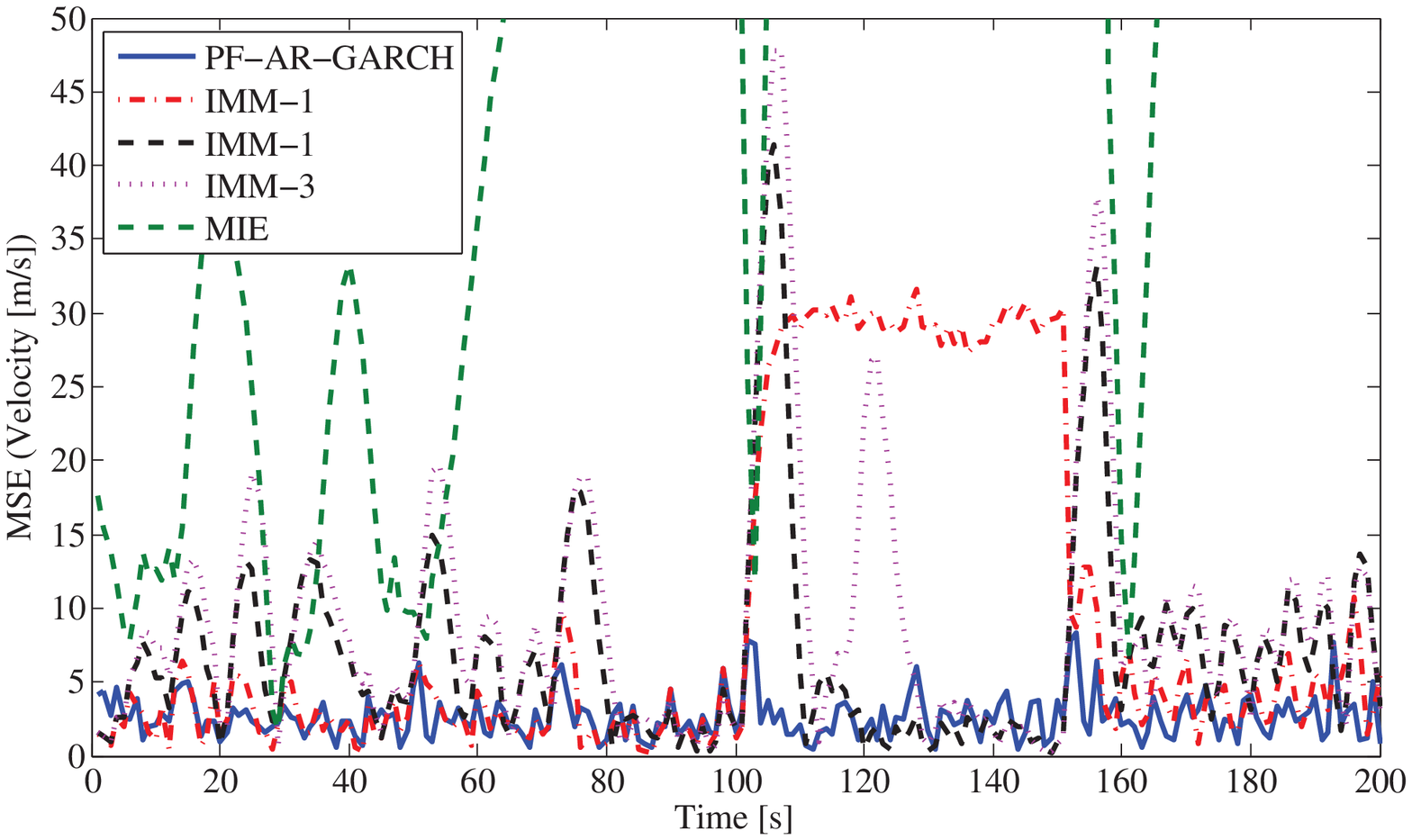}}
\centerline{(b)}
\end{minipage}
\caption{The MSE of different methods in 200 sec for (a) acceleration and (b) velocity}
\label{sinmse}
\end{figure}

For better comparison, we also evaluate MSE
of estimated acceleration by running
a simulation for the above
methods. The results are depicted in Fig. \ref{sinmse}.
However, the results of the PF approach and some time intervals of MIE algorithm are excluded from Fig. \ref{sinmse}
because they do not perform well in this scenario.


\section{Conclusion}
In this paper, we introduced a semi-parametric
category of the target tracking. The parameters
are embedded in AR-GARCH modeling of
volatility process. The stochastic volatility and target's state are estimated by a bootstrap PF, simultaneously.
We proposed maneuvering motion model
of a target using an SDE whose volatility is subsequently modeled
by a GARCH process.
%model for tracking maneuvering targets is discussed.
This approach allowed us to evaluate the performance of SDE model
based on GARCH process for tracking a maneuvering target with
abrupt changes in its acceleration.
It is shown that the GARCH
model provides finer tracking accuracy for high maneuvering target
by using heavy
tailed distribution modeling, i.e. GARCH, instead of Gaussian distribution adopted by
IMM, MIE, and etc., for system noise of state space model.
The simulation results determine that the proposed PF-AR-GARCH approach, models
and tracks abrupt changes in acceleration more accurately. Consequently,
the proposed stochastic volatility modeling enhances
maneuvering target tracking performance.

\appendices
\section{Conditional Heteroscedasticity of Acceleration}
In this appendix, we show that the stochastic property of autoregression coefficient
of acceleration, $\zeta(k)$, in (\ref{AccSinger1}) results conditional heteroscedasticity.
If we suppose that the immediate past information for acceleration is $
\Psi_{k-1}=\{a(0),a(1), \cdots, {a(k-1)}, w(0), w(1), \cdots, w(k-1)\}$, the conditional mean of
$a(k)$ will be obtained be conditional expectation of (\ref{AccSinger1}):
\begin{equation} \label{AppBconditionalmean}
E \left. \left\{a(k)\right| \Psi_{k-1} \right\}=
E \left. \left\{\zeta(k)\right| \Psi_{k-1} \right\} a(k-1)
\end{equation}
Using (\ref{AppBconditionalmean}), the conditional variance of $a(k)$ is written as:
\begin{small}
\begin{eqnarray} \label{AppBconditionalvariance}
\nonumber &&\mbox{Var} \left. \left\{a(k)\right| \Psi_{k-1} \right\} \\
&&\nonumber
=E \left. \left\{ \left(a(k)
- E \left. \left\{\zeta(k)\right| \Psi_{k-1} \right\} a(k-1)\right)^2 \right| \Psi_{k-1} \right\}
\\
&&\nonumber
=E \left. \left\{ \left(\zeta(k)a(k-1)
- E \left. \left\{\zeta(k)\right| \Psi_{k-1} \right\} a(k-1) + w(k)\right)^2 \right| \Psi_{k-1}\right\}
\\
&& =\sigma_{\zeta}^2 a^2(k-1)+\sigma^2_m.
\end{eqnarray}
\end{small}
According to (\ref{AppBconditionalvariance}), it is clear that conditional variance of
$a(k)$ is not constant in time, then, $a(k)$ is conditional heteroscedastic.

\section{Covariance Matrix Calculation}
In this appendix, we solve the stochastic It$\bar{\mbox{o}}$
integral to obtain the suggested time-varying covariance matrix
for maneuvering target tracking application proposed in this paper
which is applied in bootstrap PF procedure. To achieve this, we
begin with equation (\ref{U}). We suppose that the SV,
$\mbox{\boldmath $H$}(s)$, is constant and set to $\mbox{\boldmath $H$}(kT)$ in the integration interval
$[kT,kT+T)$,
which is estimated based on GARCH process in the filtering procedure. Thus,

\begin{small}
\begin{eqnarray}
\label{apen1} \nonumber \mbox{\boldmath $U$}(k) &=& \int_{kT}^{kT+T} \exp(\mbox{\boldmath
$F$}(kT+T-s)) \mbox{\boldmath $G$} \mbox{\boldmath $H$}^{\frac{1}{2}}(kT) d\mbox{\boldmath $W$}(s) \\
\nonumber
&=& \int_{kT}^{kT+T} \mbox{\boldmath $\Phi$}(kT+T-s,\mu) \mbox{\boldmath $G$} \mbox{\boldmath $H$}^{\frac{1}{2}}(kT) d\mbox{\boldmath $W$}(s) \\
\nonumber
&=&
\int_{kT}^{kT+T}
\left[ {\begin{array}{*{20}{c}}
{\mbox{\boldmath $\varphi$}_1(s)}\\
{\mbox{\boldmath $\varphi$}_2(s)}\\
{\exp ( - \mu (kT + T - s)) \mbox{\boldmath $I$}_{2\times2}}
\end{array}} \right] \\
&& \left[ {\begin{array}{*{20}{c}}
{\sqrt{h_x(kT)}} & {0}\\
{0}     &  {\sqrt{h_y(kT)}}
\end{array}} \right] \left[ {\begin{array}{*{20}{c}}
{dW_x(s)}\\
{dW_y(s)}
\end{array}} \right],
\end{eqnarray}
\end{small}

where $\mbox{\boldmath $\varphi_1$}(s) = \varphi_1(kT+T-s,\mu) \mbox{\boldmath $I$}_{2\times2}$
and $\mbox{\boldmath $\varphi_2$}(s) = \varphi_2(kT+T-s,\mu) \mbox{\boldmath $I$}_{2\times2}$.
Then, we easily have:
\begin{equation}
\label{aa} \mbox{\boldmath $U$}(k) =
\int_{kT}^{kT+T}
\left[ {\begin{array}{*{20}{c}}
{{\gamma _{1x}(s)}}&0\\
0&{{\gamma _{1y}(s)}}\\
{{\gamma _{2x}(s)}}&0\\
0&{{\gamma _{2y}(s)}}\\
{\gamma_{3x}(s)}&0\\
0&{\gamma_{3y}(s)}
\end{array}} \right]
\left[ {\begin{array}{*{20}{c}}
{dW_x(s)}\\
{dW_y(s)}
\end{array}} \right],
\end{equation}
where
\begin{eqnarray}
\label{qq} \nonumber \gamma _{1x}(s) &=& \varphi_1(kT+T-s,\mu) \sqrt{h_x(kT)}, \\
\nonumber \gamma _{1y}(s) &=& \varphi_1(kT+T-s,\mu) \sqrt{h_y(kT)}, \\
\nonumber \gamma _{2x}(s) &=& \varphi_2(kT+T-s,\mu) \sqrt{h_x(kT)}, \\
\nonumber \gamma _{2y}(s) &=& \varphi_2(kT+T-s,\mu) \sqrt{h_y(kT)}, \\
\nonumber \gamma _{3x}(s) &=& \exp ( - \mu (kT + T - s)) \sqrt{h_x(kT)}, \\
\nonumber \gamma _{3y}(s) &=& \exp ( - \mu (kT + T - s)) \sqrt{h_y(kT)},
\end{eqnarray}
and we define matrix $\mbox{\boldmath $\Gamma$}(s)$ as follow:
\begin{equation}
\label{Gamma} \mbox{\boldmath $\Gamma$}(s) =
\left[ {\begin{array}{*{20}{c}}
{{\gamma _{1x}(s)}}&0\\
0&{{\gamma _{1y}(s)}}\\
{{\gamma _{2x}(s)}}&0\\
0&{{\gamma _{2y}(s)}}\\
{\gamma_{3x}(s)}&0\\
0&{\gamma_{3y}(s)}
\end{array}} \right].
\end{equation}
Thus
\begin{equation}
\label{aqa} \mbox{\boldmath $U$}(k) =
\int_{kT}^{kT+T}
\mbox{\boldmath $\Gamma$}(s)
\left[ {\begin{array}{*{20}{c}}
{dW_x(s)}\\
{dW_y(s)}
\end{array}} \right].
\end{equation}
According to the input vector $\mbox{\boldmath $U$}(k)$ in equation (\ref{aqa}),
$E \left\{ {\mbox{\boldmath $U$}(k) \mbox{\boldmath $U$}^T(i)} \right\}$ can be calculated as follow:
\begin{eqnarray}
\label{martingle} \nonumber && E
\left\{ {\mbox{\boldmath $U$}(k) \mbox{\boldmath $U$}^T(i)} \right\}
= \\ \nonumber &&
E \biggr\{
\int_{kT}^{kT+T}
\mbox{\boldmath $\Gamma$}(s)
\left[ {\begin{array}{*{20}{c}}
{dW_x(s)}\\
{dW_y(s)}
\end{array}} \right]
\int_{iT}^{iT+T}
\left[ {\begin{array}{*{20}{c}}
{dW_x(\tau)}\\
{dW_y(\tau)}
\end{array}} \right]^T \mbox{\boldmath $\Gamma$}^T(\tau) \biggr\} \\
&=&
\int_{kT}^{kT+T} \!\!\!\! \int_{iT}^{iT+T}
\!\!\!\! \mbox{\boldmath $\Gamma$(s)}
E
\biggr\{ \!\!
\left[ {\begin{array}{*{20}{c}}
{dW_x(s)}\\
{dW_y(s)}
\end{array}} \right] \!\!
\left[ {\begin{array}{*{20}{c}}
{dW_x(\tau)}\\
{dW_y(\tau)}
\end{array}} \right]^T \biggr\}
\mbox{\boldmath $\Gamma$}^T(\tau).
\end{eqnarray}
According to Brownian motion properties \cite{SDE}, independent increments and
zero-mean Gaussian distribution in an increment, when $k \ne i$,
\[
\label{0} E
\biggr\{ \!\!
\left[ {\begin{array}{*{20}{c}}
{dW_x(s)}\\
{dW_y(s)}
\end{array}} \right] \!\!
\left[ {\begin{array}{*{20}{c}}
{dW_x(\tau)}\\
{dW_y(\tau)}
\end{array}} \right]^T \biggr\} = \mbox{\boldmath $0$}_{2 \times2}
\]
then, $E
\left\{ {\mbox{\boldmath $U$}(k) \mbox{\boldmath $U$}^T(i)} \right\} = \mbox{\boldmath $0$}_{2 \times2}$.
The covariance matrix $\mbox{\boldmath $Q$}(k)$
satisfies, using (\ref{martingle}),
\begin{eqnarray}
\label{covariance} \nonumber \mbox{\boldmath $Q$}(k) &=&
E \left\{ {\mbox{\boldmath $U$}(k) \mbox{\boldmath $U$}^T(k)} \right\} \\
&=& \mathop{\int \int}_{\!\!\!kT}^{KT+T} \! \mbox{\boldmath
$\Gamma$(s)} d\mbox{\boldmath $B$}(s,\tau) \mbox{\boldmath
$\Gamma$}^T(\tau)
\end{eqnarray}
where
\begin{small}
\begin{equation}
\label{dB} d\mbox{\boldmath $B$}(s,\tau) = E \biggr\{ \!\!
\left[ {\begin{array}{*{20}{c}}
{dW_x(s)dW_x(\tau)} &\quad {dW_x(s)dW_y(\tau)}\\
{dW_y(s)dW_x(\tau)} &\quad {dW_y(s)dW_y(\tau)}
\end{array}} \right] \biggr\}.
\end{equation}
\end{small}
The Brownian motions in directions $x$ and $y$ are independent
processes from each other, so that $E \{dW_x dW_y\} = 0$.
Accordingly,  $d\mbox{\boldmath $B$}(s,\tau)$ is a diagonalized
matrix and (\ref{dB}) is verified by It$\bar{\mbox{o}}$
calculation, $E \{dW_x(s) dW_x(\tau)\} = \delta(s-\tau) ds d\tau$,
and can be written as
\begin{eqnarray}
\label{Brownian} d\mbox{\boldmath $B$}(s,\tau) = [\delta(s-\tau)
ds d\tau] \mbox{\boldmath $I$}_{2 \times2}.
\end{eqnarray}
By inserting (\ref{Brownian}) in (\ref{covariance}) the covariance
matrix $\mbox{\boldmath $Q$}(k)$ required for bootstrapping of PF
is expressed as
\begin{equation}
\label{Qn} \mbox{\boldmath $Q$}(k) = \int_{kT}^{kT+T}
\mbox{\boldmath $\Gamma$}(s) \mbox{\boldmath $\Gamma$}^T(s) ds.
\end{equation}
By substituting  (\ref{Gamma}) in  (\ref{Qn}) and computing the
Riemann integral, the matrix $\mbox{\boldmath $Q$}(k)$ will be obtained.

\ifCLASSOPTIONcaptionsoff
  \newpage
\fi

\bibliographystyle{IEEEbib}
\bibliography{ref}

\end{document}